\documentclass [aps,pra,amssymb,amsmath,twocolumn, showpacs]{revtex4-1}
\usepackage[latin9]{inputenc}
\usepackage{amsmath}
\usepackage{amssymb}
\usepackage{graphicx}
\usepackage{verbatim}
\usepackage{bm}
\usepackage{color}

\usepackage{graphicx,epsfig,amsfonts,amssymb}
\usepackage{bm}
\usepackage{times}
\usepackage{lipsum}
\usepackage{verbatim}

\newcommand{\nn}{\nonumber}

\newcommand{\bv}{\mathbf v}

\newcommand{\bw}{\mathbf w}

\newcommand{\dg}{\dagger}

\newcommand{\be}{\begin{eqnarray}}
\newcommand{\ee}{\end{eqnarray}}
\newcommand{\la}{\langle}
\newcommand{\ra}{\rangle}
\newcommand{\rar}{\rightarrow}

\begin{document}

\title{Multistate Landau-Zener models with all levels crossing at one point}

\author {Fuxiang Li}
\affiliation{Center for Nonlinear Studies, Los Alamos National Laboratory, B258, Los Alamos, NM
87545}
\affiliation{Theoretical Division, Los Alamos National Laboratory, B258, Los Alamos, NM
87545}
\author {Chen Sun}
\affiliation{Department of Physics, Texas A\&M University, College Station, TX, 77843}
\affiliation{Theoretical Division, Los Alamos National Laboratory, B258, Los Alamos, NM
87545}

\author{Vladimir Y. Chernyak}
\address{Department of Chemistry and Department of Mathematics, Wayne State University, 5101 Cass Ave, Detroit, Michigan 48202, USA}
\author {Nikolai A. Sinitsyn}
\affiliation{Theoretical Division, Los Alamos National Laboratory, B258, Los Alamos, NM
87545}

\date{\today}

\begin{abstract}
 We discuss  common properties and reasons for integrability in the class of multistate Landau-Zener (MLZ) models with all diabatic levels crossing at one point.  Exploring the Stokes phenomenon, we show that each previously solved model has a dual one, whose scattering matrix can be also obtained analytically. For applications, we demonstrate how our results can be used to study conversion of molecular into atomic Bose condensates during passage  through the Feshbach resonance, and  provide purely algebraic solutions of the bowtie and
 special cases of the driven  Tavis-Cummings model (DTCM).
 \end{abstract}


\date{\today}

\maketitle

\section{Introduction}
This article continues the series of publications \cite{multiparticle,six-LZ,cQED-LZ,chen-LZ,constraints} about exact results in the MLZ theory \cite{be}. This theory deals with explicitly time-dependent Schr\"odinger equations of the form
\be
i\frac{d}{dt} \psi = \hat{H}(t) \psi, \quad \hat{H}=\hat{B}t+\hat{A},
\label{mlz}
\ee
where $\hat{A}$ and $\hat{B}$ are Hermitian time-independent $N\times N$ matrices.
An MLZ model is called {\it solvable} if one can determine the scattering matrix $\hat{S}$ or transition probability matrix $\hat{P}$, $P_{nm}\equiv |S_{nm}|^2$, for evolution from infinite negative to infinite positive time values. Recent discovery of integrability conditions \cite{six-LZ} has led  to numerous fully solvable models of this type. Complexity of these models varies from the elementary two-state case \cite{LZ,book-LZ} and few-state models  \cite{six-LZ}  to systems of spins and fermions with nonlinear interactions and combinatorially large phase space \cite{cQED-LZ,chen-LZ}.

\begin{figure}
  \scalebox{0.45}[0.45]{\includegraphics{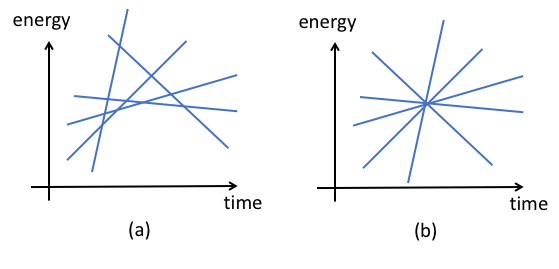}}
\hspace{-2mm}\vspace{-4mm}
\caption{(a) Diabatic level diagram of some arbitrary model of type (\ref{mlz}). (b) This diagram for a model with all levels crossing in one point.  }
\label{levels-fig}
\end{figure}
There are two reasons making the MLZ theory attractive. First, scattering matrices in solvable MLZ models are obtained explicitly in terms of the known special functions. This is in contrast to time-independent systems that are considered solvable by the algebraic Bethe ansatz.  For finite size models, Bethe ansatz generally produces only implicit solutions that depend on parameters satisfying complex nonlinear algebraic equations \cite{tv-bethe,qd-goden}, which in turn are often almost as hard to study numerically as to diagonalize the full Hamiltonian by generic algorithms.
For this reason, there are still debates about what  are the criteria for calling a finite size quantum model  integrable \cite{yuzbashyan-LZ}.
Within the MLZ theory, the notion of quantum integrability acquires conventional meaning of completely analytical description of the scattering process.

Second, MLZ theory opens unusual opportunities for
studies of mesoscopically complex quantum systems without approximations.
Many solved MLZ models describe situations that have previously been discussed in experimental context, for example, they describe interacting spin systems in linearly time-dependent magnetic fields \cite{li-16LZ}. The same models  have been used previously to explain experiments with molecular nanomagnets \cite{app-spin}, although analytical solutions of these models had not been known. Similarly, the DTCM \cite{cQED-LZ}, which belongs to the class (\ref{mlz}), had been extensively studied by variety of approximate and numerical methods
before its solution was found \cite{gurarie-LZ}. Originally, this model attracted attention due to applications to experiments with molecular Bose condensates \cite{app-bose}. More generally, multistate Landau-Zener transitions appear in numerous modern fields of research, including quantum coherence \cite{coher}, dynamic phase transitions \cite{dziarmaga},  Landau-Zener interferometry \cite{LZ-interferometry},  metrology \cite{metrology-LZ}, quantum control \cite{qcontrol}, superconductors \cite{sc-LZ}, ultracold atoms \cite{ultracold-LZ}, and  quantum dots \cite{dot-lz-exp1}.


In MLZ theory, eigenvalues of matrix $\hat{B}t$ are called diabatic levels. It is usually convenient to visualize parameters of the Hamiltonian by plotting diabatic levels in a time-energy diagram, as shown in Fig.~\ref{levels-fig}(a). Models in which all diabatic levels are crossing at one point, as shown in Fig.~\ref{levels-fig}(b), are obtained by setting diagonal elements of $\hat{A}$ to zero. Such models are particularly important in MLZ theory because they are used as elementary blocks to construct more complex solvable systems using integrability conditions \cite{Quest-LZ}.
Unfortunately, these conditions  do not apply to cases with all crossings happening at a single point. Therefore, such models need specific research on their own.

In this article, we present several findings that reveal symmetries and some of the reasons for integrability in MLZ  theory. We also develop methods and mathematical concepts that should be handy in future studies of this topic. The structure of our article is as follows. In section II we collect known solvable MLZ models featuring  all levels crossing in one point, and then discuss their bipartite property. In section III we point to another symmetry of solvable systems, and then use  these symmetries to design purely algebraic derivation of transition probabilities in the bowtie and five-state DTCM. Section IV discusses Stokes phenomenon in MLZ systems. We show that bipartite property leads to a new class  of solvable models with  applications to physics of molecular Bose condensates. We summarize our findings in section V.

\section{Bipartite Models}

\begin{figure}
  \scalebox{0.3}[0.3]{\includegraphics{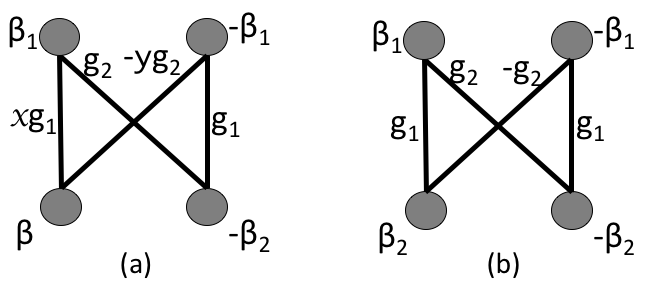}}
\hspace{-2mm}\vspace{-4mm}
\caption{Connectivity graph of the Hamiltonian (\ref{four-h}): (a) general case (b) at values of parameters $x=y=1$ and $\beta=\beta_2$. }
\label{connect-fig1}
\end{figure}

When all levels are crossing at a single point, it is convenient to illustrate parameters of the Hamiltonian as a connectivity graph, which is different from the diabatic level diagram in Fig.~\ref{levels-fig}. Let us think about the diabatic states as  the nodes of the
connectivity graph. If direct coupling between two such states, i.e., the corresponding off-diagonal element of the matrix $\hat{A}$, is  nonzero, we will connect corresponding nodes with a link. One can then mark links by corresponding couplings and nodes by corresponding level slopes, as shown in Fig.~\ref{connect-fig1}(a). This figure contains all information about  the Hamiltonian:
\be
\hat{H}(t) = \left(
\begin{array}{cccc}
\beta_1t & xg_1 & g_2 & 0 \\
xg_1 & \beta t & 0 & -yg_2 \\
g_2 & 0 & -\beta_2t& g_1 \\
0 & -yg_2 & g_1 & -\beta_1 t
\end{array}
\right),
\label{four-h}
\ee
where $g_1,\,g_2,\, x,\, y,\, \beta,\, \beta_1,\, \beta_2$ are constant parameters. It is known that this model is solvable \cite{Quest-LZ} under condition
\be
x=\sqrt{\frac{\beta_1-\beta}{\beta_1-\beta_2}}, \quad y=\sqrt{\frac{\beta_1+\beta}{\beta_1+\beta_2}}.
\label{xy1}
\ee
Interestingly, the procedure that leads to solution of this model remains mathematically unjustified, although numerous careful numerical checks have always  confirmed its validity. Only at special choice of parameters, shown in Fig.~\ref{connect-fig1}(b), has this solution been  proved both analytically and exactly \cite{constraints}.

\begin{figure}
  \scalebox{0.25}[0.25]{\includegraphics{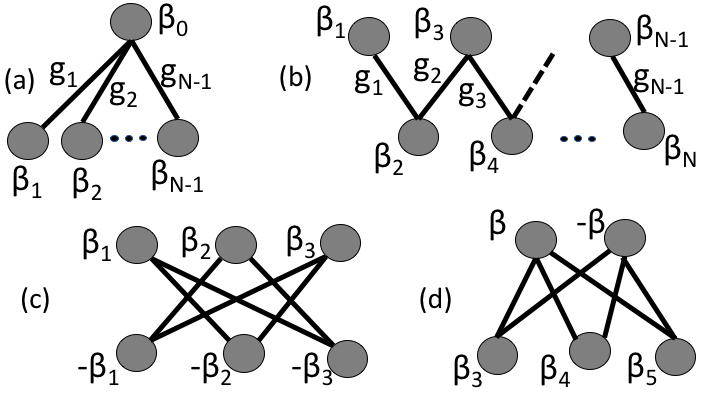}}
\hspace{-2mm}\vspace{-4mm}
\caption{Connectivity graphs of (a) bowtie model, (b) LZ-chain model, (c) composite model generated by populating four-state bowtie model with two noninteracting fermions with couplings defined in (\ref{h6b}). (d) Five-state bipartite model with couplings and solution discussed in appendix B.}
\label{connect-fig2}
\end{figure}

Let us now show, in Fig.~\ref{connect-fig2}, connectivity graphs of other known solvable models.  Figure~\ref{connect-fig2}(a) corresponds to the bowtie model \cite{bow-tie}, in which an
 arbitrary number of crossing levels are coupled directly to only one of them. Corresponding Schr\"odinger equation for amplitudes $a_n$,
$n=0,1,\ldots, N-1$ reads
\be
i\dot{a}_0=\beta_0 t a_0 +\sum_{k=1}^{N-1} g_k a_k; \,\,\, i\dot{a}_k = \beta_k t +g_ka_0, \,\,\, k\ne0.
\label{schr1}
\ee

Figure~\ref{connect-fig2}(b) corresponds to  the LZ-chain model \cite{constraints}, in which each state is coupled only to states with nearby indexes. Diabatic state amplitudes of this model evolve according to equations
\be
i\dot{a}_n=\beta_n t a_n + g_n a_{n+1} +g_{n-1} a_{n-1}, \,\,\,\, n=1,\ldots,N,
\label{schr2}
\ee
where we assume that $g_0=g_N=0$.

Not all models of the form (\ref{schr2}) have scattering matrices that are known. The most general solvable LZ-chain model is the DTCM. This model has equidistant level slopes,  $\beta_n = \beta n$, and couplings given by,
\be
g_n = g \sqrt{N_B+n}\sqrt{S(S+1)-(S-n+1)(S-n)}.
\label{TC-g}
\ee
where $g$ and $N_B>-1$ are constant parameters, and $2S+1=N$. In the limit $N_B\rar \infty$ but $g^2N_B={\rm const}$, this model describes a quantum spin of size $\hat{S}$ in a linearly time-dependent field.

General solution of the DTCM is currently known only in the form of an algorithm that allows derivation of transition probabilities in any invariant sector of this model recursively from solutions for  lower dimensional sectors \cite{cQED-LZ}. Complexity of this procedure is growing very quickly. Therefore, only special cases are well understood, such as the four-state sector ($S=3/2$), and the case of arbitrary $S$ but specific initial conditions at $t=-\infty$: with either
$|a_n|^2=\delta_{n1}$ or $|a_n|^2=\delta_{nN}$.

Any solvable MLZ model can be used to construct more complex ones by ``populating" the original model with noninteracting
fermions or bosons \cite{multiparticle,dziarmaga,jipdi}. An example of such a composite model is shown in Fig.~\ref{connect-fig2}(c). Its Hamiltonian reads
\begin{equation}
\hat{H} = \left( \begin{array}{cccccc}
\beta_1t & 0 & 0 & g_{13} & g_{12} &0 \\
0 & \beta_2t & 0 &g_{23} & 0 & -g_{12}\\
 0&0 & \beta_3t & 0 & -g_{23} & -g_{13} \\
g_{13} &g_{23}&0 &-\beta_3t & 0& 0 \\
g_{12} & 0 &-g_{23} &0 & -\beta_2t & 0 \\
0 & -g_{12} & -g_{13} & 0& 0 & -\beta_1t
\end{array} \right).
\label{h6b}
\end{equation}
We provide details of its solution in appendix A.

Finally, Fig.~\ref{connect-fig2}(d) shows the connectivity graph of the previously unknown five-state model whose solution we present in appendix B. This model was found and solved by considering a special case of a more general five-state system  that can be solved using
integrability conditions, as it was done for the four-state model (\ref{four-h}) in \cite{Quest-LZ}. There is no doubt that this class of models can be extended to higher dimensions, although their solutions quickly become very complex.

Let us now observe the most obvious common property of graphs in Figs.~\ref{connect-fig1}~and~\ref{connect-fig2}: all of the known solvable models are represented by bipartite graphs, i.e. graphs whose nodes can be partitioned into two groups so that nodes of the same group are connected only to
nodes of the other group. It is natural to expect then that new solvable MLZ systems can be found among similar bipartite systems.

In what follows, we will refer to all MLZ models with a  bipartite connectivity graph and with all diabatic level crossings at a single point in the diabatic level diagram as simply {\it bipartite models}.
So, let us explore this property and its effect on the scattering matrix.

As there are two interacting groups of states, $G_1$ and $G_2$, we give the index 1 to the smaller group of nodes and the index 2 to the bigger group in the connectivity graph (or arbitrarily if these groups have equal numbers of elements), and then
introduce operator $\hat{\Theta}$, which is represented by a diagonal $N\times N$ matrix with
\be
\Theta_{kk}=(-1)^{f_k},\quad \Theta_{ij}=0, \quad i\ne j.
\label{theta-def}
\ee
Here and in what follows, $f_k$ is the index (either 1 or 2) of the group to which the $k$-th diabatic state belongs. It is straightforward to verify that Hamiltonians of  bipartite MLZ models  satisfy the following identity:
\be
\hat{\Theta} \hat{H}(-t) \hat{\Theta} = -\hat{H}(t),
\label{id1}
\ee
from which it follows also that
\be
\hat{\Theta} e^{-i\hat{H}(-t)dt} \hat{\Theta} = e^{i\hat{H}(t)dt}.
\label{id11}
\ee

Consider the formal expression for the evolution matrix
\be
\hat{U}(T,-T) = \lim \limits_{dt \rightarrow 0} \prod_{n=-T/dt}^{T/dt} e^{-i\hat{H}(t_n)dt},
\label{ev1u}
\ee
where $t_n=ndt$ and $n$ is increasing in the product from the right to the left. We can then insert resolution of identity $\hat{1} =\hat{\Theta} \hat{\Theta}$ after each factor of the product in (\ref{ev1u}) to obtain that
\be
\label{id2}
\hat{\Theta} \hat{U}(T,-T) \hat{\Theta}=\prod_{n=-T/dt}^{T/dt} \hat{\Theta} e^{-i\hat{H}(t_n)dt} \hat{\Theta} = \\
\nonumber =\prod_{n=-T/dt}^{T/dt} e^{i\hat{H}(-t_n)dt}=\hat{U}^{\dagger}(T,-T).
\ee
Using definition of the scattering matrix \cite{constraints}, $\hat{S}= \lim \limits_{T\rar \infty} U(T,-T)$, and writing (\ref{id2}) in components, we find relation between
scattering matrix elements and between corresponding transition probabilities:
\be
S_{nm}=(-1)^{f_n+f_m} S_{mn}^*,\quad P_{nm}=P_{mn}.
\label{id3}
\ee
Thus, the transition probability matrix of any bipartite model is symmetric. Moreover, diagonal elements $S_{nn}$ of the scattering matrix are purely real.

Symmetries (\ref{id3}) have been known in bowtie and LZ-chain models  \cite{bow-tie,constraints}.
However, algebraic derivation along (\ref{id1})-(\ref{id2}) can reveal a new and less obvious than (\ref{id3}) constraint: let us take the trace of the evolution operator times $\hat{\Theta}$. Using the fact that under the trace we can perform cyclic permutations, we can rewrite
\be
{\rm Tr}\left[\hat{U}(T,-T) \hat{\Theta} \right] = {\rm Tr}\left[\hat{U}(0,-T) \hat{\Theta} \hat{U}(T,0)\right].
\label{id41}
\ee
Identity (\ref{id11}) then leads to simplifications:
\be
\nonumber e^{-i\hat{H}(T)dt} \hat{\Theta} e^{i\hat{H}(-T)dt} = e^{-i\hat{H}(T)dt} \hat{\Theta} e^{-i\hat{H}(-T)dt} \hat{\Theta} \hat{\Theta} = \\
\nonumber =e^{-i\hat{H}(T)dt} e^{i\hat{H}(T)dt}  \hat{\Theta} = \hat{\Theta}.
\ee
We can repeat this step until we get rid of all exponential factors. The result is a new identity for the scattering matrix elements:
\be
{\rm Tr} \left[ \hat{S} \hat{\Theta} \right]={\rm Tr} \left[ \hat{\Theta} \right],
\label{id4}
\ee
or in components:
\be
 \sum_{n\in G_2} S_{nn}-\sum_{m\in G_1} S_{mm} = N-2M,
\label{id42}
\ee
where $G_1$ and $G_2$ are, respectively, the first and the second groups of diabatic states;
 $N$ is the size of the Hilbert space and $M$ is the number of diabatic states in the first group.

 Although constraint (\ref{id42}) is linear in scattering matrix elements it leads to a nonlinear constraint on transition probabilities. For example, for sufficiently small couplings we can rewrite (\ref{id42}) as
\be
 \sum_{n\in G_2} \sqrt{P_{nn}}-\sum_{m\in G_1} \sqrt{P_{mm}} = N-2M.
\label{id43}
\ee
At larger couplings, we already should care about signs of scattering matrix elements, so that generally we can only say that either $S_{nn} = \sqrt{P_{nn}}$ or $S_{nn}=-\sqrt{P_{nn}}$.  The fact that simple symmetries, such as Eq.~(\ref{id1}), can lead to nonlinear constraints on transition probabilities,  such as Eq.~(\ref{id43}), is the first important finding of this article.

We conclude this section by pointing to one more consequence of symmetry (\ref{id3}). It is known that elements of the scattering matrix of any MLZ model satisfy the so-called hierarchy constraints (HCs) \cite{constraints}.
Let us assume that indexes of diabatic states are chosen so that $\beta_1>\beta_2>\ldots>\beta_N$. Then the two lowest order HCs read:
\begin{eqnarray}
\label{be1}
\noindent S_{11}&=& e^{- \pi \sum \limits_{k=2}^N |A_{k1}|^2/(\beta_1-\beta_k)}, \\
\nonumber \\
\label{hhhc2}
\noindent  {\rm Det} \left(
\begin{array}{cc}
S_{11} & S_{12} \\
S_{21} & S_{22}
\end{array}
\right) &=&  e^{- \pi \sum \limits_{k=3}^{N} \left( \frac{| A_{k1}|^2}{(\beta_1-\beta_k)}+\frac{ |A_{k2}|^2}{(\beta_2-\beta_k)} \right)},
\end{eqnarray}
\begin{widetext}
and generally:
\be
{\rm Det} \left(
\begin{array}{cccc}
S_{11}& S_{12}& \cdots & S_{1M} \\
S_{21} & S_{22} & \cdots &S_{2M} \\
\vdots & \cdots  &  \ddots & \vdots \\
S_{M1} & \cdots & \cdots & S_{MM}
\end{array}
\right) =  e^{-\pi \sum \limits_{k=M+1}^{N} \sum \limits_{r=1}^{M} \frac{|A_{kr}|^2}{(\beta_r-\beta_k)} },\quad M =1,\ldots, N-1.
\label{h222}
\ee
\end{widetext}

In an arbitrary MLZ model, the first constraint (\ref{be1}) can be converted into expression for the probability to remain on the same level if the system is initially  on the level with the extremal slope.
The second constraint (\ref{hhhc2}) already depends on phases of amplitudes $S_{22}$, $S_{12}$, and $S_{21}$, so Eq.~(\ref{hhhc2}) cannot generally lead to a constraint that is written only in terms of transition probabilities.
However, for bipartite models we know that $S_{22}$ is real, and from (\ref{id3}) we find
\begin{widetext}
\be
S_{11}S_{22}-(-1)^{f_1+f_2} |S_{12}|^2 = e^{-\pi \left( \sum \limits_{k\in \bar{G}(1)} |A_{1k}|^2/(\beta_1-\beta_k)+\sum \limits_{k\in \bar{G}(2)} |A_{2k}|^2/(\beta_2-\beta_k)\right) },
\label{hc-cons1}
\ee
 where $\bar{G}(k)$ means the set of elements of the group of states connected to the node $k$ directly. Using that $|S_{12}|^2\equiv P_{12}$ and (\ref{be1}),
we find relation between probabilities $P_{22}$ and $P_{12}$:
\be
e^{-2\pi  \sum \limits_{k\in \bar{G}(1)} |A_{1k}|^2/(\beta_1-\beta_k)} P_{22} =
\left( (-1)^{f_1+f_2} P_{12}+e^{-\pi \left( \sum \limits_{k\in \bar{G}(1)} |A_{1k}|^2/(\beta_1-\beta_k)+\sum \limits_{k\in \bar{G}(2)} |A_{2k}|^2/(\beta_2-\beta_k)\right) } \right)^2.
\label{hc-cons2}
\ee
\end{widetext}
Finally, we note that HCs can be reformulated so that they start not with the highest but rather with the lowest slope level. Such HCs lead to equations analogous to (\ref{be1})-(\ref{hc-cons2}), in which indexes $1,2,\ldots,N$ are replaced by, respectively, $N,N-1,\ldots,1$. This is followed then by change of slope signs, e.g., $\beta_1-\beta_2 \rar \beta_{N-1}-\beta_N$, e.t.c.. So, in any bipartite model,  not only survival probabilities for two extremal levels are known but also there are two equations of the form (\ref{hc-cons2}) that impose coupling-dependent relation between $P_{12}$ and $P_{22}$, as well as between $P_{N,N-1}$ and $P_{N-1,N-1}$.

\section{Pure gauge phase ansatz}
The class of bipartite models is considerably larger than the class of solved models shown in Figs.~\ref{connect-fig1} and \ref{connect-fig2}. What else is special to solvable cases?

It has been noticed previously \cite{six-LZ,Quest-LZ,gbow-tie} that many MLZ models can be solved by  a semiclassical ansatz  in which  direct transition to another level brings a simple phase $i$ to the transition amplitude.
This suggests the following ansatz to satisfy constraints (\ref{id3}):
\be
S_{mn}=\alpha_{mn} i^{f_n+f_m} e^{i(\phi_{m}-\phi_{n})},
\label{sym-spec1}
\ee
 where $\phi_{m}$ are  phases that may depend on parameters of the model, possibly nontrivially, and where $\alpha_{mn}$ are real parameters. We will think of $\alpha_{mn}$ as elements of a new matrix $\hat{\alpha}$. This ansatz reduces the number of unknown phases from $N(N-1)/2$ to just $N$.
Symmetry (\ref{sym-spec1}) can be verified analytically in  several cases: the  bowtie model, the models in Figs.~\ref{connect-fig1}(b) and \ref{connect-fig2}(c), and the model of arbitrary spin in a linearly time-dependent magnetic field,  because their scattering matrices are known \cite{bow-tie,jipdi,four-LZ}.


However,  in the DTCM and  the general case of the model (\ref{four-h}), only transition probabilities have been found so far, while phases of scattering matrix elements remain unknown. So, we cannot verify (\ref{sym-spec1}) in them analytically. Instead, we tested for applicability of the ansatz (\ref{sym-spec1}) numerically. Since phases $\phi_m$ are unknown, we did it indirectly, namely, Eq.~(\ref{sym-spec1}) predicts that any cyclic product of amplitudes should not depend on all phases $\phi_m$. So, if we define
\be
\quad c_3\equiv S_{13} S_{32}S_{21}, \quad c_4 \equiv S_{12}S_{24} S_{43}S_{31}.
\label{cdef}
\ee
 and set indexes so that $1,3 \in G_1$ and $2,4 \in G_2$, then
\be
{\rm Im}[c_3] ={\rm Im}[c_4]=0.
\label{ntest1}
\ee
We found that, indeed, scattering matrix elements in {\it all} previously solved bipartite models satisfy (\ref{ntest1}).
\begin{figure}
  \scalebox{0.30}[0.30]{\includegraphics{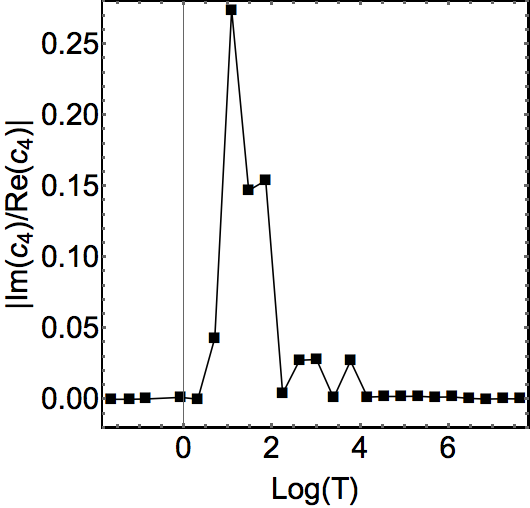}}
\hspace{-2mm}\vspace{-4mm}
\caption{Numerically obtained ratio of imaginary and real parts of the product $c_4(T) =  U_{12}(T,-T)U_{24}(T,-T) U_{43}(T,-T)U_{31}(T,-T) $ as function of $T$ in model (\ref{four-h}). Simulation algorithm is explained in Ref.~\cite{cQED-LZ}. Parameters:
$\beta=0.1$, $b_2=0.65$, $b_1=1.25$, $g_1=0.37$, $g_2=0.5$; $x,\, y$  satisfy condition (\ref{xy1}). Discrete dots are results of numerical simulations and the solid line is the guide for eyes.
Note the logarithmic scale of the time axis. The last point corresponds to $T_{\rm max}=2038$ and ${\rm Im} [c_4(T_{\rm max})]/{\rm Re}[c_4(T_{\rm max})]=2.9\times 10^{-4}$. Analogous tests were performed for DTCM sectors with up to $N=6$ and different values of parameters $N_B$ and $g$. Results showed  similar behavior  of $c_{3,4}$ (not shown), confirming validity of Eq.~(\ref{sym-spec1}).}
\label{test-sym}
\end{figure}
There is  difference from Eq.~(\ref{id3}), however, in that Eq.~(\ref{sym-spec1}) generally does not work if evolution operator $\hat{U}(T,-T)$ at finite $T$ is substituted instead of the scattering matrix. So, if we define $c_3(T)$ and $c_4(T)$ by replacing elements of $\hat{S}$ by elements of $\hat{U}(T,-T)$ and plot the ratios $r_{3,4}={\rm Im} \left(c_{3,4}(T)\right)/{\rm Re} \left( c_{3,4} (T) \right)$ as functions of $T$, we find that these ratios are initially noticeably nonzero but decrease to zero at increasing $T$, as we show in Fig.~\ref{test-sym}. Hence, the ansatz (\ref{ntest1}) works only asymptotically. We found this behavior also in the bowtie and driven Tavis-Cummings models. There are also exceptions, where (\ref{sym-spec1}) holds for evolution operator $c_4(T)$ at arbitrary $T$, e.g., in the model in Fig.~\ref{connect-fig1}(b). We attribute this to presence of additional symmetry in this model, which was discussed in \cite{four-LZ}.

It is expected that known solvable models do not exhaust the solvable class of bipartite systems. If so, conditions (\ref{sym-spec1}) are likely common or even determining this class. Their understanding may shed new light on the phenomenon of integrability in MLZ theory and provide the way to generate new solvable cases.
Therefore, below we explore some of the consequences of  this symmetry.

\subsection{Numerical search for models with symmetry (\ref{sym-spec1}) }

If MLZ-integrability is indeed related to the symmetry (\ref{sym-spec1}),
then conditions (\ref{ntest1}) provide a strategy to search for new solvable models.
Consider for example a four-state LZ-chain:
\be
\hat{H}= \left(
\begin{array}{cccc}
\beta_1t &g_1 &0 &0 \\
g_1 & \beta_2t &g_2 &0\\
0& g_2 &\beta_3t & g_3 \\
0&0&g_3 & \beta_4t
\end{array}
\right).
\label{c-four}
\ee
The most general known solvable model of the type (\ref{c-four}) is the four-state DTCM that has equally distant slopes $\beta_n=\beta n$ and couplings constants specified by Eq.~(\ref{TC-g}).

\begin{figure}
  \scalebox{0.30}[0.30]{\includegraphics{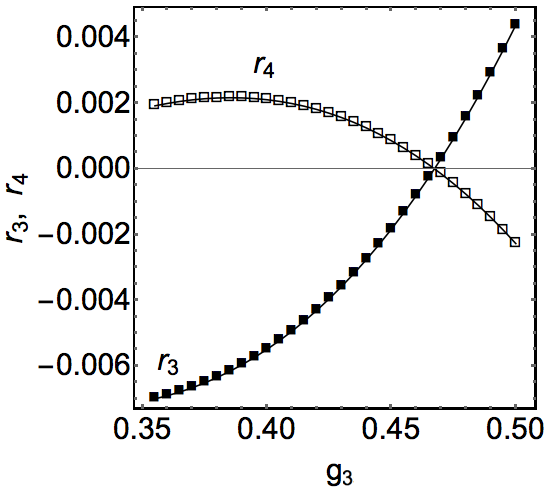}}
\hspace{-2mm}\vspace{-4mm}
\caption{Numerically obtained ratios  $r_{3,4} = {\rm Im}(c_{3,4})/{\rm Re}(c_{3,4})$ in the model (\ref{c-four}). Parameters:
$\beta_4=0$, $\beta_3=1$, $\beta_2=2$, $\beta_1=5$. $g_1=g_2=0.5$. Numerical evolution is from $-T$ to $T$, at $T=250$. Discrete dots are results of  simulations and the solid lines are the guide for eyes.}
\label{test-deform}
\end{figure}

 Let us  consider some case of Eq~(\ref{c-four}) with a different, from DTCM,  choice of slopes and also with fixed values of two out of the three couplings.
We can then simulate evolution with the Hamiltonian (\ref{c-four}) and check the validity of (\ref{ntest1}) at different values of the variable coupling. We show results of such a test in Fig.~\ref{test-deform}. Although constant parameters are chosen differently from the DTCM,
this figure shows that there is a coupling value $g_3\approx 0.47$, at which imaginary parts of both $c_3$ and $c_4$ change their signs simultaneously. So, at this point, symmetry (\ref{sym-spec1}) is valid and the model is likely solvable although its solution is currently not known.

Several similar tests  correctly reproduced coupling values in a few already solved models (not shown).
We also found that  such points are rather common in bipartite models with four interacting states. However, in models with higher dimensions, scanning of a single parameter is usually insufficient to  observe such  points.
We also found that some symmetric choices of parameters, for example LZ-chains with equal couplings and equidistant level slopes, do not generally satisfy (\ref{ntest1}) for $N\ge4$. So, our numerical investigation suggests that there are numerous solvable but still unknown bipartite models with asymmetric choices of parameters.

\subsection{Parametrization by orthogonal symmetric matrix}

Let us now note that $N$ phases  $\phi_{n}$ in (\ref{sym-spec1}) can be removed by mere change of phases in definition of $N$ diabatic states followed by redefinition of phases of coupling constants. 
These phases drop from all expressions for HCs because determinants depend only on cyclic products such as in (\ref{cdef}).
Therefore, each of $N-1$ independent HCs leads to an equation on elements of $\hat{\alpha}$, and consequently on transition probabilities. This is in contrast to the general bipartite case in which only four such constraints are allowed.

Second, from (\ref{id3}) it follows that $\hat{\alpha}$ has to be symmetric:
\be
\alpha_{mn}=\alpha_{nm},
\label{sym2}
\ee
and from unitarity, $\hat{S}\hat{S}^{\dag}=\hat{1}$, we find that $\hat{\alpha}$ is orthogonal. Together, Eqs.~(\ref{sym2}) and (\ref{sym3}) mean that
\be
\hat{\alpha}^2=\hat{1}.
\label{sym3}
\ee
An immediate consequence is that  eigenvalues of $\hat{\alpha}$ are all $\pm 1$. Moreover, we can find multiplicities of such eigenvalues. At zero couplings we have $S_{nm} = \delta_{nm}$ and
 eigenvalues of $\hat{\alpha}$ coinciding with its diagonal elements:
\be
\alpha_{nn}=(-1)^{f_n}, \quad {\rm for} \quad \hat{A}=0.
\label{ev1}
\ee
Switching on couplings continuously cannot change discrete eigenvalues. So, we find that the number of $-1$ eigenvalues of $\hat{\alpha}$ equals the number $M$ of nodes in the smaller group, $G_1$.

Symmetric orthogonal matrix with $M\le N/2$ negative eigenvalues can be parametrized with components of $M$ orthonormal vectors:
\be
\hat{\alpha}=\hat{1}-2\sum_{{\bf k}=1}^M {\bf v_k}^T {\bf v_k}.
\label{num1}
\ee
Indeed, ${\bf v_k}$ is the eigenvector of $\hat{\alpha}$ with eigenvalue $-1$. All real vectors orthogonal to ${\bf v}_1 \ldots {\bf v}_M$ belong to the space with
eigenvalues $+1$.

Vector ${\bf v_1}$ depends on $N-1$ independent parameters (the last one is fixed by requirement of its unit norm).
Vector ${\bf v_2}$ depends on $N-2$ unknown parameters because, in addition to normalization, it must be orthogonal to vector ${\bf v}_1$.
We can continue this counting until we reach ${\bf v_M}$ that depends on $N-M$ independent parameters. This corresponds totally to $M(2N-M-1)/2$ unknown parameters of the matrix $\hat{\alpha}$.
On the other hand, we have only $N-1$ HCs that express combinations of these parameters via the elements of the model's Hamiltonian.

It is interesting to know whether knowledge of the symmetry (\ref{sym-spec1}) supplemented by HCs, and possibly the constraint (\ref{id42}), is sufficient to derive all transition probabilities in some models.
For example, the bowtie model has $M=1$. The number of independent parameters in $\hat{\alpha}$ is then $N-1$, which is the same as the number of independent HCs.
The next in complexity case with $M=2$ shows that HCs alone are not sufficient to reconstruct all parameters of  $\hat{\alpha}$. For example, for $N=4$ we have then
five parameters for $\hat{\alpha}$ and only three independent HCs. In the case of the driven Tavis-Cummings model, however, we have additional information because all transition probabilities from the extremal slope levels are known when $N$ is arbitrary \cite{cQED-LZ}. In any case, HCs are highly nonlinear, so it is not obvious that they can be disentangled to reconstruct elements of the scattering matrix.
In the following two subsections we will show that in some cases this can be done.



\subsection{Algebraic solution of the bowtie model}

Previously, solution of the bowtie model has been obtained by methods of complex analysis. It is expected, however, that there is a hidden algebraic reason behind any exact solution.
Below we claim that, in the bowtie model, the reason is in  combination of HCs (\ref{h222}) and the property (\ref{sym-spec1}).

For a general case of the bowtie model, it is convenient to choose state indexes so that
 $\beta_i >\beta_j$ if $i>j$, and  the special level has the slope $\beta_0$. So, we allow negative integer indexes.  Let there be $m$ diabatic level slopes lower than $\beta_0$ and $n$ higher slopes  ($m + n + 1=N$). Nonzero elements of the Hamiltonian are:
\be
\label{elH}
H_{j0}=H_{0j} = g_j,\quad H_{jj}=\beta_jt, \quad j=-m,\ldots, n.
\ee

According to (\ref{num1}), matrix $\hat{\alpha}$ can be expressed via only one real vector
$$\bv = (v_{-m}, \ldots, v_{-1}, v_0, v_1, \ldots, v_{n}),$$
 with a constraint $\sum_{j } v_j^2 =1$.
Moreover, it is easy to verify that transition probabilities in this case must depend only on combinations $2v_k^2$, which we are going to determine now.

Let us consider the first $m$ equations of HCs (\ref{h222}) for the Hamiltonian (\ref{elH}). In the left hand sides, they will have the determinant
\be
  {\rm Det}(1-2 \bw \bw^T),
\label{det1}
\ee
where, for $(j+m+1)$-st  hierarchy equation, vector $\bw$ is obtained by  keeping only components of $\bv$ up to $j$-th index (note that $j<0$).
A determinant  written as (\ref{det1}) can be simplified using Sylvester's identity \cite{sylvester}:
$${\rm Det} (\hat{I}_n+\hat{R}\hat{Q}) = {\rm Det} (\hat{I}_m+\hat{Q}\hat{R}),
$$
where $\hat{R}$ is $m\times n$ and $\hat{Q}$ is $n \times m$ matrices. Applying this formula to (\ref{det1}) we find
\be
 {\rm Det}(\hat{1}-2 \bw \bw^T)=(1-2\bw^T \bw )= 1-2\sum_{k\le j} v_k^2.
 \label{silvest}
 \ee
Thus, first $m$ HCs  can be written as
\be
\label{bt-id1}
1-2 \sum_{k \le j} v_k^2 = \prod_{k\le j} p_k, {~\rm for ~} -m \le j<0,
\ee
where we defined
\be
p_j = e^{- \pi g_j^2 /|\beta_0 -\beta_j|}, \quad j\ne 0.
\ee
Solving Eqs.~(\ref{bt-id1}) recursively starting from $j=-m$ we find
\be
&&2 v_j^2 = (1- p_j) \prod_{k<j} p_k,  {~\rm for~~} j<0.
\ee
Similarly,  we can consider the HCs for $n\times n$ part in the right-lower corner of the scattering matrix. They  lead to $n$ constraints:
\be
&&2 v_j^2 = (1-p_j) \prod_{k>j} p_k, {~\rm for~~ } j>0.
\ee
 The component $v_0$ of the vector $\bv$ is obtained by requiring the unit norm of $\bv$:
 \be
 \label{be-id0}
 2 v_0^2 = \prod_{j>0} p_j + \prod_{j<0} p_j.
 \ee

Substituting (\ref{bt-id1})-(\ref{be-id0}) into (\ref{num1}) we find the scattering matrix up to $N$ phases $\phi_k$, which can be gauged away. Taking the absolute value squared of components of this matrix we obtain transition probabilities, which indeed coincide with the ones known for the bowtie model. For example,
\be
P_{0\rar0} = (1-2v_0^2)^2=\left(1-\prod_{j>0} p_j - \prod_{j<0} p_j \right)^2.
\label{bt00}
\ee
Such a solution of the bowtie model is considerably simpler than the original one in \cite{bow-tie}.

\subsection{Driven Tavis-Cummings model}
\begin{figure}
  \scalebox{0.35}[0.35]{\includegraphics{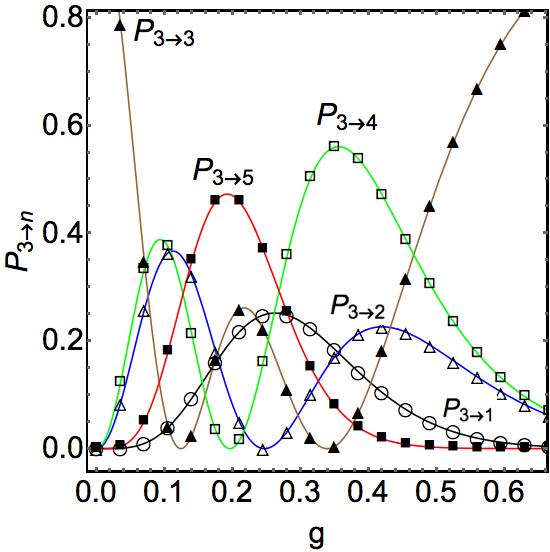}}
\hspace{-2mm}\vspace{-4mm}
\caption{Transition probabilities from the third level to all diabatic states of the five-state DTCM at different values of the coupling. Solid curves are predictions of Eqs.~(\ref{p-tc1})-(\ref{p-tc11})
and discrete points are results of numerically calculated absolute value squared  elements of the evolution matrix,
$P_{3\rar n} = |U_{n3}(T,-T)|^2$, at $T=2000$ and $N_B=0$. The time step for simulations is $dt=0.00001$. Details of the numerical algorithm can be found in \cite{cQED-LZ}. }
\label{tc5test}
\end{figure}
DTCM is a LZ-chain model with equidistant slopes $\beta_n=\beta n$ and couplings given in (\ref{TC-g}). Alternatively, its Hamiltonian describes interaction of a bosonic mode with an arbitrary size spin:
\be
\hat{H}=t \hat{a}^{\dg}\hat{a} +g\left( \hat{a}^{\dg} \hat{S}^{-}+\hat{a}\hat{S}^+ \right),
\label{TC-H}
\ee
where we used the fact that we can always set $\beta=1$ by rescaling time, as discussed in \cite{cQED-LZ}. Implicitly, the model (\ref{TC-H}) depends on the additional parameter $N_B$, which is the number of bosons in the state with the fully up-polarized  spin.
It is known how to derive transition probabilities in this model by taking some limit of the known solution of a much more complex model \cite{cQED-LZ}. The latter, however, has combinatorially large phase space, so this process
 quickly becomes too involved. The exception is for conditions when the spin is initially fully polarized either down or up, i.e., when we start on the level with, respectively, highest or lowest slopes. Then, according to \cite{chen-LZ}, transition probabilities to other states are described by q-deformed binomial distributions.
For example, let us introduce parameters
\be
\label{defs1}
\nonumber a \equiv e^{-2\pi g^{2}N_B}, \quad x \equiv  e^{-2\pi g^2}, \quad p_{n} \equiv a x^n, \\
 q_n \equiv 1-p_n, \quad N\equiv 2S+1.
\ee
We know then from \cite{cQED-LZ} that
\be
\nonumber |S_{11}|^2&=& p_1^{2S} = (ax)^{2S}, \,\,\, |S_{N,N}|^2= p_{2S}^{2S} = (ax^{2S})^{2S},\\
|S_{12}|^2 &=& q_1(p_1^{2S-1}+ \\
\nonumber &+&p_1^{2S-2}p_2 +p_1^{2S-3}p_2^2 +...+ p_2^{2S-1}),
\label{s12-1}
\ee
\be
|S_{N,N-1}|^2  &=& q_{2S} (p_{2S}^{2S-1}+p_{2S}^{2S-2}p_{2S-1} + \\
\nonumber &+&p_{2S}^{2S-3}p_{2S-1}^2 + \ldots + p_{2S-1}^{2S-1}),\\
\nonumber |S_{1N}|^2&=&q_1q_2\ldots q_{2S}.
\label{s12-1}
\ee

Let us rewrite these elements in terms of $x$ and $N_B$:
$$
|S_{12}|^2 = \frac{x^{(N_B+1)(2S-1)}(1-x^{N_B+1})(1-x^{2S})}{1-x}.
$$
$$
|S_{N,N-1}|^2 =   \frac{x^{(N_B+2S-1)(2S-1)}(1-x^{N_B+2S})(1-x^{2S})}{1-x}.
$$
In what follows, to simplify expressions, we will consider only the case with $N_B=0$. Then
\be
\label{scat2}
\nonumber |S_{12}|^2 = x^{2S-1}(1-x^{2S}), \\
|S_{N,N-1}|^2  = x^{(2S-1)^2}(1-x^{2S})^2/(1-x), \\
\nonumber |S_{1N}|=(1-x)(1-x^2)\ldots (1-x^{2S}).
\ee
At $N_B=0$ we have couplings
\be
\nonumber g_1^2&=& g^2 (2S), \quad g_2^2=2g^2(4S-2), \quad g_3^2=18g^2(S-1), \\
g_{2S}^2&=&g^2 (2S)^2, \quad g_{2S-1}^2=g^2 2(2S-1)^2.
\ee
Equations~(\ref{hc-cons1})  and (\ref{scat2}) allow us to derive $S_{22}$ and $S_{N-1,N-1}$ explicitly for any DTCM chain:
\be
\label{s22-tc}
S_{22}&=&x^{3S-2}-x^{S-1}(1-x^{2S}), \\ 
\label{s22s-tc}
S_{N-1,N-1}&=&x^{1+2S(S-2)}\left(1-\frac{(x^{2s}-1)^2}{1-x} \right).
\ee

For $N=4$, i.e., $S=3/2$,
this  is sufficient to reconstruct all elements of the transition probability matrix  using the fact that this matrix is doubly stochastic and that for bipartite models it is
symmetric:
\begin{widetext}
\be
\hat{P}^{(S=3/2)}=
\ee
\be
\nonumber \left( \begin{array}{cccc}
x^3& x^2(1-x^3) & x(1+x)(1-x)^2(1+x+x^2) & (1-x)^3(1+x) (1+x+x^2) \\
x^2(1-x^3) &x(x^3+x^2-1)^2 & (x^2-1)(x^3+x^2+x-1)^2 & x(1+x)(1-x^3)^2 \\
x(1+x)(1-x)^2(1+x+x^2) &(x^2-1)(x^3+x^2+x-1)^2  & x(x(x^3+x^2+x-1)-1)^2 &x^4(1-x)(1+x+x^2)^2 \\
(1-x)^3(1+x) (1+x+x^2) & x(1+x)(1-x^3)^2 & x^4(1-x)(1+x+x^2)^2 & x^9
\end{array} \right).
\label{s4}
\ee
\end{widetext}
We checked that this solution coincides with the one derived in \cite{cQED-LZ}.

Let us now look at previously unexplored case with $N=5$, i.e., $S=2$.
The following elements of the transition probability matrix are known from \cite{cQED-LZ}:
\be
\label{p-tc1}
\nonumber && P_{11}=x^4, \quad P_{12}=P_{21} = x^3 (1 - x^4), \\
\nonumber &&P_{13}=P_{31} = x^2 (1-x^3)(1-x^4), \\
&& P_{14} =P_{41} =x(1-x^2) (1-x^3)(1-x^4), \\
\nonumber &&P_{15} =P_{51} =(1-x)(1-x^2) (1-x^3)(1-x^4), \\
\nonumber &&P_{25}=P_{52} = x(1-x^2) (1-x^3) (1-x^4) (1+x + x^2 + x^3), \\
\nonumber &&P_{35} =P_{53} = x^4(1-x^3)(1-x^4)(1+x^2) (1+x+x^2) ,\\
\nonumber &&P_{45}=P_{54} = x^9 (1-x^4)(1+x+x^2 + x^3), \quad P_{55} = x^{16}.
\ee
Using the 2nd order HCs, we have
\be
S_{22} = x (-1 + x^3 + x^4),
\ee
\be
S_{44} = x^2 (-1 - x - x^2 + x^3 + x^4 + x^5 + x^6).
\ee
The trace formula (\ref{id42}) gives
 $$
 S_{11} - S_{22} + S_{33} - S_{44} + S_{55} =1,
 $$
 from where we find
\be
S_{33} = 1 - x - 2 x^2 - x^3 + 2 x^5 + x^6 + x^7.
\ee
The remaining three elements $P_{23}$, $P_{24}$, $P_{34}$ can be again obtained using the fact that summation of elements of each column or each row of the transition probability matrix is equal to one.
We found:
\begin{widetext}
\be
\label{p-tc11}
\nonumber &&P_{23} = P_{32} =x - 2 x^3 - 3 x^4 - x^5 + 4 x^6 + 5 x^7 + 3 x^8 - x^9 - 3 x^{10} -  2 x^{11} - x^{12}, \\
&&P_{24} = P_{42} = 1 - 2 x - 2 x^2 + x^3 + 4 x^4 + 6 x^5 - 4 x^7 - 6 x^8 - 4 x^9 + x^{10} + 2 x^{11} + 2 x^{12} + x^{13}, \\
\nonumber &&P_{34} = P_{43} = x + 2 x^2 - 4 x^4 - 7 x^5 - 4 x^6 + 3 x^7 + 8 x^8 + 9 x^9 +  4 x^{10} - x^{11} - 4 x^{12} - 4 x^{13} - 2 x^{14} - x^{15}.
\ee
\end{widetext}
Figure~\ref{tc5test} shows  perfect agreement of Eqs.~(\ref{p-tc1})-(\ref{p-tc11}) with results of numerical simulations.

The ansatz (\ref{sym-spec1}) was not needed to solve $N=4$ and $N=5$ sectors. Rather we used Eqs.~(\ref{id3}), (\ref{id42}), and (\ref{hc-cons1}) that are common to all bipartite models.
For the higher dimensional DTCM sector with $N=6$, however, counting free parameters shows that even with additional input about phases and transition probabilities from fully polarized states, we cannot reconstruct all elements of the transition probability matrix.
Apparently, unlike the bowtie model, simple discrete symmetries plus HCs are already not sufficient to obtain complete analytical understanding of DTCM.  So, it seems that validity of the ansatz (\ref{sym-spec1}) is important but this is  not the last piece of the puzzle about the origin of LZ-integrability.

\section{Stokes phenomenon in MLZ models}

\subsection{General case}
Evolution equation with an MLZ Hamiltonian corresponds to a system of ordinary differential equations with a single irregular point in a complex plane at $t\rightarrow \infty$.
It has been discussed previously \cite{no-go,shytov,joye-LZ} that since the MLZ Hamiltonian has no singular points at finite complex time values, it is possible to deform the time-evolution contour so that it connects real $t\rar \pm \infty$ limits with a contour that goes along a semicircle with radius $R\rar \infty$. Deformations of time contour in both upper and lower halves of the complex plane are allowed.

The advantage of this deformation is that if the integration contour always has $|t| \rar \infty$ then the diabatic levels are always well separated, and the eigenvectors of the Hamiltonian always coincide with diabatic states.
 Adiabatic
approximation predicts the amplitudes of solutions
\be
\psi_n(t) = e^{-i\int_{t_0}^{t} \epsilon_n(t) \, dt}+O(1/t), \quad n=1,\ldots, N,
\label{adiab}
\ee
where  $\epsilon_n$ is the adiabatic energy of the $n$-th eigenstate of the Hamiltonian.
Moreover, one does not need to know exact expressions for eigenvalues and eigenvectors because only nonvanishing in $|t|\rar \infty$ limit terms contribute substantially to Eq.~(\ref{adiab}). Thus, eigenvectors coincide with
diabatic states and, for MLZ models with all levels crossing at one point, we can approximate
\be
\epsilon_n(t) \approx \beta_n t+\sum_{m \ne n} \frac{|g_{nm}|^2}{(\beta_n-\beta_m)t}.
\label{enmlz1}
\ee
For definiteness, we will assume that $t_0$ in (\ref{adiab}) is taken at infinite negative real values.

\begin{figure}
  \scalebox{0.15}[0.15]{\includegraphics{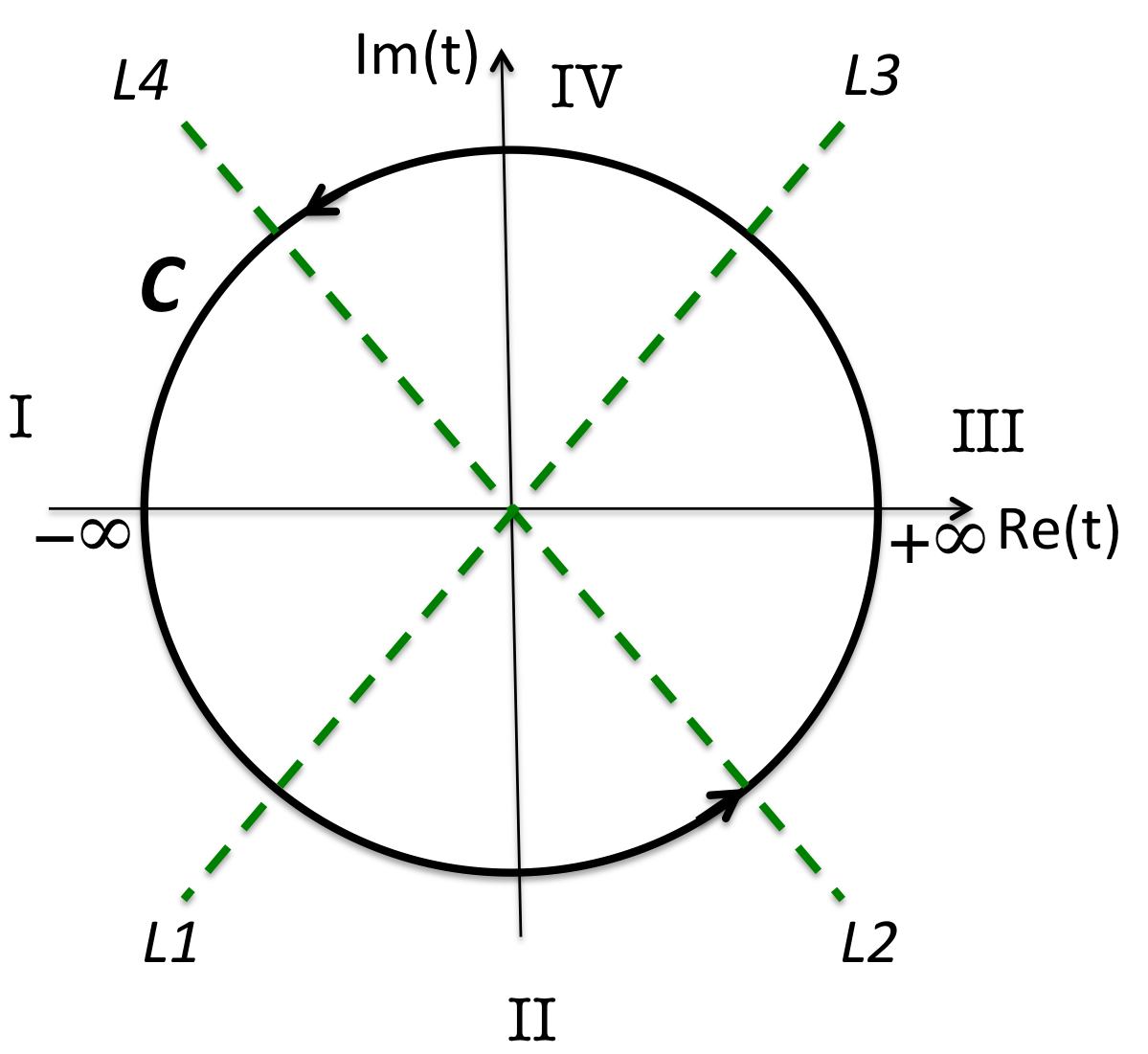}}
\hspace{-2mm}\vspace{-4mm}
\caption{Contour ${\bm C}$ in complex time plane with $|t|\rar \infty$.  Stokes lines $L_1$-$L_4$ (dashed green) mark directions of highest positive or negative ramp rates of asymptotic solutions.
Four anti-Stokes lines correspond to negative and positive parts of real and imaginary time axes.
Four Stokes sectors are marked by I, II, III, and IV. Each of them, by definition, encloses space with one and only one anti-Stokes line. So, in principle, neighboring Stokes sectors overlap in the vicinity of Stokes lines.
 Arrows show the choice of the positive direction (counterclockwise) of time evolution along the contour ${\bm C}$.
 Scattering matrix is obtained by following contour ${\bm C}$ in the lower half of the complex plane and crossing two Stokes lines, $L_1$ and $L_2$, from sector I to sector III. }
\label{b1-fig}
\end{figure}

Asymptotic functions (\ref{adiab}), however, cannot describe the complete solution along the contour that goes near the  irregular point (at infinity in our case), which is the essence of the Stokes phenomenon \cite{math,aoki-LZ}. For example,
the true solution must be a single-valued function of $t$ but, upon integrating along the full circle ${\bm C}$,  Eq.~(\ref{adiab}) acquires the factor:
$
\psi_n(te^{2\pi i}) =e^{2\pi \eta_{n}} \psi_n(t),
$
 $n=1,\ldots, N$,
where we introduced 
\be
\hat{\eta} = {\rm diag}\{ \eta_1, \eta_1, \ldots, \eta_N \}, \,\,\, \eta_n = \sum \limits_{m \ne n} \frac{|g_{nm}|^2}{(\beta_n-\beta_m)}.
\label{etas1}
\ee
So, solution cannot behave as in Eq.~(\ref{adiab}) everywhere along the contour ${\bm C}$.
Let us define anti-Stokes lines  \cite{note} along which solutions (\ref{adiab}) do not grow or decay exponentially in the limit $|t| \rar \infty$.
Stokes sectors are defined then as parts of the complex plane that contain one and only one anti-Stokes line, as we additionally explain in Fig.~\ref{b1-fig}.  It is an established mathematical fact that it is always possible to find a set of $N$
solutions with asymptotic behavior (\ref{adiab}) everywhere within such a sector \cite{deform-book}.

The problem is that knowledge of the leading term of the solution in one Stokes sector is not sufficient to find asymptotic behavior of the solution in the full
neighboring sector. What can be shown, however, is that such a continuation is described by transition matrices associated with the Stokes lines \cite{note}.  Along these lines, the functions (\ref{adiab}) have the highest or lowest
rates of the amplitude growth when approaching the singular point at infinity.


Figure~\ref{b1-fig} shows  that there are four Stokes lines in MLZ models  given by $t=Re^{i(-3\pi/4+(k-1)\pi/2)}$, where $R\in (0,+\infty)$, and $k=1,2,3,4$, and there are four anti-Stokes lines that are negative and positive parts of the real and imaginary axes. Thus, the covering of the complex plane
by Stokes sectors  is relatively simple: there are only four of them, marked by $I-IV$. Note that neighboring sectors overlap along Stokes lines.

Let us now assume that at $t=-\infty$ on the real time axis the state vector is a sum of solutions (\ref{adiab}) taken with some amplitudes $a_n^I$.  In sector-I   this solution can be analytically continued, and  along the contour ${\bm C}$ with
infinite radius we have
\be
\psi(t)= \sum_{n=1}^N a_n^I \psi^I_n(t),
\label{psi-I}
\ee
where $ \psi^I_n(t)$ are solutions of the full Schr\"odinger equation that have asymptotics (\ref{adiab}) in the whole sector-I.
Our goal is to find the scattering matrix for evolution from $t=-\infty$ to $t=+\infty$ on the real time axis, which is essentially the same as to find how the function $\psi(t)$ defined in (\ref{psi-I}) behaves near the point of crossing of the positive real axis with the
contour ${\bm C}$, which is inside sector-III. To get there along the contour ${\bm C}$, we should first find solution in  the intermediate sector-II.

Solutions with asymptotic behavior (\ref{adiab}) that can be analytically continued throughout the whole sector-II
have different subleading contributions from functions $\psi^I_n(t)$ at the same time values in the region where the two sectors overlap. Therefore, the vector amplitudes in overlapping sectors must be related:
\be
{\bm a}^{II}=\hat{S}_1 {\bm a}^I,
\label{psi-2}
\ee
and in sector-II, the state vector is given by
\be
\psi(t) = \sum_{n=1}^N a_n^{II} \psi^{II}_n(t),
\label{psi-II}
\ee
where $ \psi^{II}_n(t)$ are solutions of the Schr\"odinger equation that have leading asymptotic values (\ref{adiab}) everywhere inside sector-II.

More generally, after the time-integration contour crosses a $k$-th Stokes line, we can switch to the basis of states that have leading behavior (\ref{adiab}) everywhere inside $(k+1)$-st Stokes sector.
Amplitudes in the previous and new bases become connected by some transition matrix $\hat{S}_k$. So, $\psi(t)$ defined in (\ref{psi-I}) can be rewritten as
\be
\label{stm1}
{\bm a}^{k+1} &=& \hat{S}_{k}\ldots \hat{S}_1{\bm a}^I, \\
\nonumber \psi(t) &=& \sum_{n=1}^N a_n^{(k+1)} \psi^{(k+1)}_n(t), \quad k=1,2,\ldots .
\ee
In (\ref{stm1}) and later we will identify indexes $I,II,III,...$ with $1,2,3,...$.

Matrices $\hat{S}_k$ are nontrivial, i.e.,  there is no known  general way to write them in a closed form for an arbitrary ordinary differential equation with time-dependent coefficients, including  for
 high order equations  that describe MLZ models.
However, several properties of these matrices can be rigorously proved.

 First, from the fact that solutions   $ \psi^{k}_n(t)$ and $\psi^{(k+1)}_n(t)$ in the neighboring sectors can be simultaneously analytically continued
in the overlapping region between sectors $k$ and $(k+1)$, it follows that matrix  $\hat{S}_k$ is time independent. Second, since
 such two solutions with the same leading exponents along the Stokes line are  different only by asymptotically subdominant contributions in the overlapping region,
diagonal components of $\hat{S}_k$ are all units.
Third, one can show \cite{berry-stokes} that inside the $k$-th Stokes sector if one moves along the contour ${\bm C}$ then new sub-leading exponents in solutions $\psi_{n}^k(t)$ appear sharply
 in the vicinity of the $k$-th Stokes line.
Therefore, further along this contour, amplitudes of terms with smaller ramp rate along the Stokes line will be influenced by amplitudes of terms with the higher rates but not vice versa.
Hence, matrices  $\hat{S}_k$ have a triangular form, i.e., they have all zero entries either above or below the main diagonal when state indexes follow the order of the level slopes.
 More rigorous proof of this fact can be found in  \cite{deform-book}.

In our MLZ models, along  $L_1$, we have leading asymptotics of diabatic states
$$
\psi_n^{L_1} \sim e^{\beta_n |t|^2/2}.
$$
Since we assume that $\beta_1>\ldots>\beta_N$, we have that, along $L_1$, the state with index $n$ dominates over states with higher indexes $n'>n$,
so the corresponding transition matrix  has the following triangular form:
\be
\hat{S}_1 = \left(
\begin{array}{ccccc}
1 & 0 &0 & \cdots & 0\\
x_{21} & 1 & 0 &\cdots &0\\
x_{31}&x_{32}&1 & \cdots &0\\
\vdots & \vdots & \vdots &\ddots & \vdots \\
x_{N1} & \cdots & \cdots & x_{N,N-1} &1
\end{array}
\right),
\label{stokes1}
\ee
with some constant parameters $x_{ij}$, $i,j=1,\ldots, N$.
Along Stokes line $L_2$, basis states have opposite ramp rate order, i.e.,
$$
\psi_n^{L_2} \sim e^{-\beta_n |t|^2/2},
$$
so the transition matrix is again triangular but has  nonzero elements in the upper angle:
\be
\hat{S}_2 = \left(
\begin{array}{ccccc}
1 & x_{12} & x_{13} & \cdots & x_{1N}\\
0& 1 & x_{23} &\cdots & x_{2N}\\
\vdots & \vdots &\ddots & \vdots & \vdots \\
0 & \cdots & 0 &\ddots & x_{N-1,N} \\
0 &  \cdots & 0&0 &1
\end{array}
\right).
\label{stokes4}
\ee
Analogously, we have:
\be
\label{stokes32}
\hat{S}_3 = \left(
\begin{array}{ccccc}
1 & 0 &0 & \cdots & 0\\
y_{21} & 1 & 0 &\cdots &0\\
y_{31}&y_{32}&1 & \cdots &0\\
\vdots & \vdots & \vdots &\ddots & \vdots \\
y_{N1} & \cdots & \cdots & y_{N,N-1} &1
\end{array}
\right), \\
\label{stokes41}
 \hat{S}_4 = \left(
\begin{array}{ccccc}
1 & y_{12} & y_{13} & \cdots & y_{1N}\\
0& 1 & y_{23} &\cdots & y_{2N}\\
\vdots & \vdots &\ddots & \vdots & \vdots \\
0 & \cdots & 0 &\ddots & y_{N-1,N} \\
0 &  \cdots & 0&0 &1
\end{array}
\right),
\ee
with some parameters $y_{ij}$, $i,j=1,\ldots, N$.

Note also that if we go around the closed contour ${\bm C}$, we should obtain the original state, so $\psi(t)_{t=-\infty} = \sum_{n=1}^N a_n^V\psi_n^V(t) =  \sum_{n=1}^N a_n^I\psi_n^I(t)$. Using
Eq.~(\ref{stm1}) we find
\be
\hat{1}=\hat{S}_4\hat{S}_3\hat{S}_2\hat{S}_1e^{2\pi \hat{\eta}},
\label{scs2}
\ee
where the last factor on the right hand side follows from the integral in (\ref{adiab}) over the full cycle ${\bm C}$.

\subsection{Scattering matrix}

Similarly to  Eq.~(\ref{scs2}), the desired scattering matrix for transition from large negative to large positive real time values is given by
\be
\hat{S}=\hat{S}_2\hat{S}_1 e^{\pi \hat{\eta}} .
\label{scs1}
\ee

 From Eq.~(\ref{scs1}), we find that the number of unknown parameters on the right hand side is lower than the number of entries of the scattering matrix $\hat{S}$.
 So, if the scattering matrix is known, we can
reconstruct Stokes matrices $\hat{S}_k$. For example,  comparing the last columns and rows in both sides of Eq.~(\ref{scs1}) we have
\be
x_{Nj}= S_{Nj}  e^{-\pi \eta_j}, \quad x_{jN}=S_{jN} e^{-\pi \eta_N},
\label{stokes2}
\ee
where $j=1, \ldots ,N-1$.
Comparing the next to the last columns and rows and using (\ref{stokes2}) we can find parameters $x_{N-1,j}$ and $x_{j,N-1}$, and so on. At each step of such calculations we would
 have to solve only linear
equations although final dependence of parameters $x_{ij}$ on $S_{ij}$ would be nonlinear.
So, the first  feature of Stokes phenomenon in MLZ systems is that knowledge of the scattering matrix determines all independent elements of the Stokes matrices.
This is not typical for systems of ordinary differential equations, and follows from relative simplicity of covering of the complex plane  by Stokes sectors in MLZ models.

The second interesting property follows from the fact that the number of independent parameters $x_{ij}$ is even lower than the number of elements of the scattering matrix.
There are $N$ remaining equations for diagonal elements of matrices in both sides of Eq.~(\ref{scs1}). These equations produce nontrivial constraints
on scattering amplitudes, i.e., on the physically important characteristics. For example, the first pair of such constraints reads
\be
\label{hc1}
S_{NN}&=& e^{\pi \eta_{N}}, \\
 \nonumber  S_{N-1,N-1} &=& e^{\pi \eta_{N-1}} + S_{N-1,N}S_{N,N-1}  e^{-\pi \eta_{N}}.
\ee
Comparing Eqs.~(\ref{hc1})~and~(\ref{h222}), we find that Eqs.~(\ref{hc1}) are nothing but the HCs, which were derived for MLZ theory in \cite{constraints} by a
different  approach (expression for $S_{NN}$ has been known for much longer time and is often called the Brundobler-Elser formula \cite{be}).

Although we merely reproduced already known Eq.~(\ref{h222}), its derivation in this section by exploring the Stokes phenomenon has useful consequences. First, so far, our discussion
of transition matrices $\hat{S}_k$ has not used unitarity of evolution along the real time axis. Therefore, formulas such as (\ref{hc1}) are valid even if the Hamiltonian is not Hermitian.
Second, unlike the derivation in \cite{constraints}, it is now clear that there are no other similar nontrivial constraints in MLZ theory, unless we impose additional symmetries on the model.

\subsection{Case of bipartite crossing at one point}
Let us now consider the unitary evolution of a bipartite system of levels crossing at one point. From (\ref{id3}) and  (\ref{scs2})-(\ref{scs1}) we get
\be
\hat{\Theta} \hat{S} \hat{\Theta} e^{-\pi \hat{\eta}}=\hat{S}_4\hat{S}_3.
\label{scs3}
\ee
or
\be
\hat{S}=e^{\pi \hat{\eta}} \hat{\Theta} \hat{S}_4\hat{S}_3\hat{\Theta} .
\label{scs4}
\ee

As we discussed in the previous subsection, there is a unique way to relate matrices $\hat{S}_{3}$, $\hat{S}_4$ with $\hat{S}$, and consequently with  $\hat{S}_1$ and $\hat{S}_2$.
Comparing (\ref{scs4}) with (\ref{scs1}) we find
\be
\hat{S}_3=e^{-\pi \hat{\eta}} \hat{\Theta} \hat{S}_1 \hat{\Theta} e^{\pi \hat{\eta}}, \quad \hat{S}_4=e^{-\pi \hat{\eta}} \hat{\Theta} \hat{S}_2 \hat{\Theta} e^{\pi \hat{\eta}},
\label{scs5}
\ee
or in components
\be
y_{ij}=(-1)^{f_i+f_j}e^{\pi (\eta_j - \eta_i)}x_{ij}.
\label{scs6}
\ee

As an example, let us look at the 3-state bowtie model with $\beta_1>\beta_2>0$:
\be
i\frac{d}{dt} \psi = \hat{H}^{bt}(t) \psi, \quad \hat{H}^{bt}=\left(\begin{array}{ccc}
\beta_1t &0 & g_1 \\
0 & \beta_2 t & g_2 \\
 g_1 & g_2 & 0
\end{array}
\right).
\label{bt3}
\ee
Evolution with this Hamiltonian is described by the scattering matrix \cite{hioe,bow-tie}
\begin{widetext}
\be
\hat{S}^{bt}=\left(\begin{array}{ccc}
s_{11} & s_{12} & s_{13} \\
s_{12}^* & s_{22} & s_{23} \\
 -s_{13}^* & -s_{23}^* & s_{33}
\end{array}
\right)=\left(\begin{array}{ccc}
p_1 & -\sqrt{p_1(1-p_1)(1-p_2)}&i \sqrt{(1-p_1)(1+p_1p_2)} \\
-\sqrt{p_1(1-p_1)(1-p_2)}& 1-p_1+p_1p_2 &   i \sqrt{p_1(1-p_2)(1+p_1p_2)} \\
i \sqrt{(1-p_1)(1+p_1p_2)} & i \sqrt{p_1(1-p_2)(1+p_1p_2)}  & p_1p_2
\end{array}
\right),
\label{bt3}
\ee
\end{widetext}
where we denoted
\be
p_1\equiv e^{-\frac{\pi |g_1|^2}{\beta_1}}, \quad p_2\equiv e^{-\frac{\pi |g_2|^2}{\beta_2}}.
\label{p1p2}
\ee
In the first equality in (\ref{bt3}) we used
the symmetry (\ref{id3}) of bipartite models, and in the second equality we wrote the known explicit solution, in which we assumed that
the gauge for diabatic states is chosen to remove nontrivial phases of scattering amplitudes.
Matrix $\hat{\eta}$ has components
\be
\hat{\eta} = {\rm diag} \left(\frac{|g_1|^2}{\beta_1},\frac{|g_2|^2}{\beta_2},-\frac{|g_1|^2}{\beta_1}-\frac{|g_2|^2}{\beta_2}   \right).
\label{j1}
\ee
Substituting (\ref{bt3}) and (\ref{j1}) into (\ref{scs1}) we find equations for $x_{ij}$, which are straightforward to solve:
\be
\label{bt3-s1}
\hat{S}_1&=&\left(\begin{array}{ccc}
1 & 0 & 0 \\
s_{12}^*p_1+s_{23}s_{13}^* /p_2 & 1 & 0 \\
 -s_{13}^* p_1& -s_{23}^* p_2 & 1
\end{array}
\right), \\
\label{bt3-s11}
\hat{S}_2&=&\left(\begin{array}{ccc}
1 &s_{12} p_2+s_{13}s_{23}^* /p_1 & s_{13}/(p_1p_2)\\
0& 1 & s_{23}/(p_1p_2) \\
0 & 0 & 1
\end{array}
\right),
\ee
and from (\ref{scs5}) we have
\be
\label{bt3-s2}
\hat{S}_3&=&\left(\begin{array}{ccc}
1 & 0 & 0 \\
s_{12}^*p_2+s_{23}s_{13}^* /p_1 & 1 & 0 \\
 -s_{13}^*/(p_1 p_2)& -s_{23}^*/(p_1 p_2)  & 1
\end{array}
\right), \\
\label{bt3-s22}
\hat{S}_4&=&\left(\begin{array}{ccc}
1 &s_{12} p_1+s_{13}s_{23}^* /p_2 & -p_1 s_{13}\\
0& 1 & -p_2 s_{23} \\
0 & 0 & 1
\end{array}
\right).
\ee

\subsection{Dual semi-infinite MLZ models}
As an application of the above formalism, we will show now that with any  $N$-state
bipartite model we can identify a dual  to it MLZ model  that describes
interactions of $N-1$ semi-infinite Landau-Zener chains. If  one of these models is solvable then we automatically know solution of its dual one.

First, we note that the scattering matrix between oscillatory non-decaying asymptotic functions is well defined in MLZ models because both negative and positive
pieces of the real time axis coincide with anti-Stokes lines. What is also special about MLZ models whose diabatic levels cross in one point is that
there is another, namely the imaginary time, axis with the same property. Therefore, we can similarly define the scattering matrix $\hat{S}'$ for evolution
from $t=-i\infty$ to $t=i\infty$ along the imaginary time.

Consider evolution  (\ref{mlz}) along  the imaginary time axis. It is convenient then to return to real time by replacing $t=i\tau$, where $\tau \in (-\infty,+\infty)$ is real. We find then
\be
i\frac{d}{d\tau} \psi =\hat{H}' (\tau)\psi, \quad \hat{H}'\equiv -\hat{B}\tau +i\hat{A},
\label{nonun1}
\ee
where we again assume that $\hat{B}={\rm diag}(\beta_1,\ldots, \beta_N)$ is nondegenerate diagonal and $\hat{A}$ is a Hermitian matrix with zero diagonal elements.

Imagine that at $\tau \rar -\infty$ the state vector $\psi(\tau)$ is normalized, $|\psi(\tau)_{\tau \rar -\infty}| = 1$, and  that its oscillating part behaves as
$ \psi(\tau)_{\tau \rar -\infty} \sim e^{i\beta_n t^2}$. The element of the scattering matrix, $S_{mn}'$, is then defined to be the time-independent prefactor of the
amplitude of the solution term that oscillates as $\sim e^{i\beta_m \tau^2}$ at $\tau \rar +\infty$.

Let us assume that matrix $\hat{A}$ describes bipartite couplings, i.e., that diabatic states can be split into two groups, $G_1$ and $G_2$,
 so that states of one group are coupled directly only to states of the other group. Then, according to previous discussion in this section, we can determine transition matrices
through Stokes lines in terms of scattering matrices of the evolution with MLZ Hamiltonian $\hat{H}=\hat{B}t+\hat{A}$.

In terms of transition matrices, scattering matrix for evolution (\ref{nonun1}) is given by
\be
\hat{S}' = e^{\pi \hat{\eta}/2} \hat{S}_3 \hat{S}_2 e^{\pi \hat{\eta}/2}.
\label{ss1}
\ee
To see this, we note that if $t_0$ in (\ref{adiab}) were chosen to be on the negative imaginary axis, we would be able to repeat previous arguments leading to the formula
analogous to (\ref{scs1}). However, changing $t_0$ to be real and negative results in additional factors: $\hat{S}_{k} \rar  e^{\pi \hat{\eta}/2} \hat{S}_k e^{-\pi \hat{\eta}/2}$, which account for
the difference between (\ref{ss1}) and (\ref{scs1}).

Thus, we conclude that knowledge of the solution of some bipartite MLZ model automatically results in knowledge of the scattering matrix of the model (\ref{nonun1}). The latter, however,
describes non-Hermitian evolution, which is formally not an MLZ model. Nevertheless, consider the following
secondary quantized quadratic Hamiltonian of $N$ interacting bosonic modes:
\be
\label{Hchain}
\nonumber  \hat{H}_B = -\sum_{k\in G_1} \beta_k t \hat{a}^{\dg}_k \hat{a}_k + \sum_{s\in G_2} \beta_s t \hat{b}^{\dg}_s \hat{b}_s +\\
+ \sum_{k\in G_1}\sum_{s\in G_2}  \left( i A_{ks} \hat{a}_k^{\dg} \hat{b}_s^{\dg} -i  A_{ks}^* \hat{a}_k \hat{b}_s \right).
\ee
This Hamiltonian is Hermitian, and in the Fock space of population states $|n_1,\ldots, n_N\ra$ it has the form (\ref{mlz}). Unitary evolution with $\hat{H}_B$ conserves
the difference between the total number of bosons between the two groups.
So the phase space splits into invariant subspaces that are different by the value of this conserved number. States in each subspace can be marked by $N-1$ discrete indexes.

 Let us introduce the following vector of operators
\be
\hat{\bm \phi} \equiv \left(
\begin{array}{c}
\hat{a}_1 \\
\vdots \\
\hat{a}_M \\
\hat{b}_1^{\dg}\\
\vdots\\
\hat{b}_{N-M}^{\dg}
\end{array}\right),
\label{phiop}
\ee
and consider its evolution with the Hamiltonian (\ref{Hchain}) in the Heisenberg picture:
\be
i\frac{d}{dt} \hat{\bm \phi} = [\hat{\bm \phi}, \hat{H}_B].
\label{evh1}
\ee
This equation is equivalent to
\be
i\frac{d}{dt} \hat{\bm \phi} = \hat{H}' \hat{\bm \phi},
\label{evh2}
\ee
where $\hat{H}'$ has the same form as defined in Eq.~(\ref{nonun1}). This observation means that if the scattering matrix $\hat{S}'$ is known for evolution equation (\ref{nonun1}), we
automatically know asymptotic values of operators in the Heisenberg picture of the model (\ref{Hchain}). For example,
\be
\hat{a}_n(+\infty) = \sum_{k\in G_1} S'_{nk} \hat{a}_k + \sum_{s\in G_2} S'_{ns} \hat{b}^{\dg}_s ,
\label{aab}
\ee
where we identify $\hat{a}_k$ and $\hat{b}^{\dg}_s$ with, respectively,  $\hat{a}_k(-\infty)$ and $\hat{b}^{\dg}_s(-\infty)$.
From such expressions we can find arbitrary correlators of operators in the final state. For example, let us assume that the initial state is characterized by well defined number of bosons of each atomic type.
 The final population of  the bosonic mode $n \in G_1$ is then given by
\be
\label{af}
\la \hat{a}^{\dg}_n (+\infty) \hat{a}_n (+\infty) \ra=  \\
\nonumber =\sum \limits_{k\in G_1}  \la \hat{a}^{\dg}_k \hat{a}_k \ra  |S'_{nk}|^2+  \sum \limits_{s\in G_2}  \la \hat{b}_s \hat{b}^{\dg}_s \ra |S'_{ns}|^2 ,
\ee
where the averaging is over the initial state at $t \rar - \infty$, so cross-correlators average to zero. Knowledge of such correlators can be then used to derive transition probabilities between any pair of diabatic states
of the Hamiltonian (\ref{Hchain}), as it was done in \cite{gurarie-LZ} for the simplest case with $N=2$.

Summarizing, any bipartite MLZ model of the form (\ref{mlz}), has a dual counterpart with the Hamiltonian (\ref{Hchain}). The latter is also an MLZ model but with the infinite number of interacting
states. Scattering matrices in dual models are related by equations (\ref{scs1}) and (\ref{ss1}).  Namely, assuming knowledge of the scattering matrix in any one of these equations, we can derive  triangular matrices $\hat{S}_k$, $k=1,2,3,4$ given by (\ref{stokes1}),
(\ref{stokes4}) and (\ref{stokes32}), and then obtain the scattering matrix from the other equation.

\subsection{Coherent conversions between molecular and atomic condensates}

Models of the type (\ref{Hchain}) have actually been encountered in physics in relation to atomic condensate creation,  which is induced by applying a linearly time-dependent magnetic field,  from a molecular condensate  during the passage through the Feshbach resonance
\cite{gurarie-LZ,yuzbashyan-LZ,kayali-LZ}. The number of molecules in the molecular condensate is assumed macroscopic so that this number is absorbed in the definition
of coupling constants in the mean field approximation.
 Coupling terms in (\ref{Hchain}) correspond then to quantum coherent chemical reaction channels:
$$
[AB] \rightarrow A_k+B_s,
$$
meaning that molecules $[AB]$ can split into pairs of atoms $A$ and $B$. Single atoms may have internal states marked by indexes $k$ and $s$. When the external field is far from the resonance, this reaction is suppressed.
 By   sweeping the magnetic field linearly throughout the resonance, we can control the number of single atoms that we produce under condition that this sweep is much faster than the lifetime of the molecular condensate.
Internal states can be different localized modes in a trap that interact differently with the linearly time-dependent external magnetic field. This leads to differences in slope parameters in (\ref{Hchain}).
So far, only simplest cases without internal states of the same atoms have been studied \cite{yuzbashyan-LZ,kayali-LZ}. Hence, below we will use
duality to explore the effect of competition between different channels of this quantum coherent chemical reaction.

For demonstration, we consider the case when the first atomic field has no internal states interacting with the external field while the second atom has two such states. The Hamiltonian of our model  is
\be
\hat{H}_B^{(3)} =  \beta_1 t \hat{b}^{\dg}_1 \hat{b}_1+ \beta_2 t \hat{b}^{\dg}_2 \hat{b}_2+ \\
\nonumber +\left( ig_1 \hat{a}^{\dg} \hat{b}_1^{\dg} + i  g_2 \hat{a}^{\dg} \hat{b}_2^{\dg} + {\rm h. c} \right),
\label{Hchain2}
\ee
where we will assume that $\beta_1>\beta_2>0$.

The corresponding dual model is the three-state bowtie model, whose scattering matrix is given in (\ref{bt3}) and transition Stokes matrices are given in Eqs.~(\ref{bt3-s1})-(\ref{bt3-s22}). Substituting them in (\ref{ss1}) we find
\begin{widetext}
\be
\hat{S}'=\left(\begin{array}{ccc}
p_1^{-1} & p_1^{-1} \sqrt{\frac{(1-p_1)(1-p_2)}{p_2}} & ip_1^{-1}\sqrt{\frac{(1-p_1)(1+p_1p_2)}{p_2}} \\
p_1^{-1} \sqrt{\frac{(1-p_1)(1-p_2)}{p_2}} &(p_1p_2)^{-1} -(1-p_1)p_1^{-1} & i (p_1p_2)^{-1}\sqrt{(1-p_2)(1+p_1p_2)}  \\
- ip_1^{-1}\sqrt{\frac{(1-p_1)(1+p_1p_2)}{p_2}}&  -i (p_1p_2)^{-1}\sqrt{(1-p_2)(1+p_1p_2)}  &(p_1p_2)^{-1}
\end{array}
\right).
\label{bt3p}
\ee
Let us now look at the simplest and experimentally most relevant case when all atomic modes are initially unpopulated.
Final average numbers of atoms in each of the three atomic types then read:
\be
\label{populations}
\nonumber \la \hat{a}^{\dg} (+\infty) \hat{a} (+\infty) \ra&=& |S'_{31}|^2+ |S'_{32}|^2=e^{2\pi \left(|g_1|^2/\beta_1 +|g_2|^2/\beta_2 \right)}-1,\\
 \la \hat{b}_1^{\dg} (+\infty) \hat{b}_1(+\infty) \ra &=& |S'_{13}|^2 = e^{\pi \left(2|g_1|^2/\beta_1 +|g_2|^2/\beta_2 \right)}\left(1-e^{-\pi |g_1|^2/\beta_1 }\right) \left(1+e^{-\pi \left(|g_1|^2/\beta_1 +|g_2|^2/\beta_2 \right)} \right),\\
\nonumber \la \hat{b}_2^{\dg} (+\infty) \hat{b}_2(+\infty) \ra &=& |S'_{23}|^2 = e^{2\pi \left(|g_1|^2/\beta_1 +|g_2|^2/\beta_2 \right)}\left(1-e^{-\pi |g_2|^2/\beta_2 }\right) \left(1+e^{-\pi \left(|g_1|^2/\beta_1 +|g_2|^2/\beta_2 \right)} \right).
\ee
\end{widetext}
These equations show that  new reaction channels make multiplicative effect
 on the number of produced single atoms and that relative importance of different reaction channels is exponentially sensitive to  couplings and the chemical potential crossing rates.

\section{Discussion}
We considered the special but important subclass of MLZ models with all diabatic levels crossing in one point. Our goal was to broaden the scope of exact results by exploring specific features that are common to all known solvable models.
We identified several such features.

First, it is the bipartite structure of all solvable models, i.e., all of them describe interactions between two groups of states, such that direct couplings between diabatic states within the same group are zero. We found that this simple symmetry  leads to Eqs.~(\ref{id3}), (\ref{id42}), and (\ref{hc-cons1}) that are not encountered in other MLZ models and   constrain elements of the
scattering matrix considerably.

 Our second finding is the additional symmetry (\ref{sym-spec1}), which is rather the ansatz with  phases of scattering amplitudes chosen to be removable by gauge transformations. Both analytical and numerical studies showed that this ansatz worked for all known solvable models.

One consequence of the validity of this ansatz is that the scattering matrix can be parametrized by a set of orthonormal vectors with real entries. In  known solvable models, the latter are always polynomials of some
parameter combinations. So, we can speculate that this property relates MLZ-integrability to the theory of orthogonal polynomials, whose importance for mathematical physics has been well recognized \cite{aborodin}. Moreover,  all independent hierarchy constraints (\ref{h222}), which generally tell very little about the
transition probability matrix, now lead to direct constraints on it. In the bowtie model, these constraints are sufficient to derive the complete transition probability matrix. However, in the  case of the driven Tavis-Cummings model we achieved only limited success by solving two low-dimensional sectors. Hence, there should be other symmetries that are to be discovered in order to  understand the integrability of the DTCM.

Finally, we showed that the Stokes phenomenon in all MLZ models has several simplifying properties that allow determination of all nontrivial Stokes multipliers from the scattering matrix. For bipartite models this leads to duality with a new solvable class of MLZ systems.
Models of this class describe interactions of molecular and atomic Bose condensates in time-dependent fields.

Our results definitely do not exhaust all interesting properties of bipartite models. It seems to us that future progress will be achieved by understanding how and why various discrete and dynamic symmetries constrain Stokes multipliers in MLZ systems.
Achieving this goal will enable discoveries of new classes of explicitly time-dependent solvable systems. 


\appendix
\section{Composite solvable model with particle-hole symmetry}
First, we note that (\ref{h6b}), is related to the Hamiltonian that describes interaction of three fermionic modes:
\be
\label{hph1}
\nonumber \hat{H}^f = \sum \limits_{n=1}^3 \beta_n t \hat{c}^{\dg}_n \hat{c}_n +  \\
+\left(g_{12} \hat{c}^{\dg}_1 \hat{c}^{\dg}_2
+g_{13} \hat{c}^{\dg}_1 \hat{c}^{\dg}_3 +g_{23} \hat{c}^{\dg}_2 \hat{c}^{\dg}_3 +{\rm h.c.}   \right).
\ee
We will assume that  $\beta_1>\beta_2>\beta_3>0$. For evolution with $\hat{H}^f$,
Heisenberg equations for annihilation and creation operators  read:
\be
\label{heis1}
\nonumber i\frac{d}{dt} \hat{c}_n&=&\beta_n t \hat{c}_n +\sum_{m=1;\, m\ne n}^3 (-1)^{\eta_{nm}} g_{nm} \hat{c}^{\dg}_m, \\
i\frac{d}{dt} \hat{c}_n^{\dg}&=&-\beta_n t \hat{c}_n^{\dg} -\sum_{m=1;\, m\ne n}^3 (-1)^{\eta_{nm}} g_{nm} \hat{c}_m,
\ee
where $\eta_{nm}=1$ if $m<n$ and  $\eta_{nm}=0$ if $m>n$.
Equations (\ref{heis1}) for operators are the same as equations for amplitudes in the model (\ref{h6b}), so if we know
the scattering matrix $\hat{S}$ for the model (\ref{h6b}) we can connect
also values of operators at $t= \pm \infty$:
\be
\left( \begin{array}{c}
\hat{c}_1(+\infty)\\
\hat{c}_2(+\infty)\\
\hat{c}_3(+\infty)\\
\hat{c}_3^{\dg} (+\infty)\\
\hat{c}_2^{\dg}(+\infty)\\
\hat{c}_1^{\dg}(+\infty)
\end{array} \right) = \hat{S}
\left( \begin{array}{c}
\hat{c}_1\\
\hat{c}_2\\
\hat{c}_3\\
\hat{c}_3^{\dg} \\
\hat{c}_2^{\dg}\\
\hat{c}_1^{\dg}
\end{array} \right),
\label{heis2}
\ee
where we identify $\hat{c}_i \equiv \hat{c}_i(-\infty)$, e.t.c.. Due to the particle-hole symmetry, evolution equations (\ref{heis1})  satisfy conditions of the No-Scattering theorem in \cite{li-16LZ}, according to which elements of the scattering matrix  $\hat{S}$ in (\ref{heis2}) along the second main diagonal are all zero:
$$
S_{16}=S_{25}=S_{34}=S_{43}=S_{52}=S_{61}=0.
$$

Next, consider evolution with the Hamiltonian (\ref{hph1}) in the sector that contains the vacuum state.
Other orthogonal states in this sector  include:
\be
|12\ra \equiv \hat{c}_1^{\dg} \hat{c}_2^{\dg} |0\ra, \,\,\, |13\ra \equiv \hat{c}_1^{\dg} \hat{c}_3^{\dg}|0\ra, \,\,\, |23\ra \equiv \hat{c}_2^{\dg} \hat{c}_3^{\dg} |0\ra.
\label{basis1}
\ee
In the basis (\ref{basis1}) the Hamiltonian (\ref{hph1}) reads:
\begin{equation}
\hat{H}_4^{bt} = \left( \begin{array}{cccc}
0 & g_{12} & g_{13} & g_{23} \\
g_{12} & (\beta_1+\beta_2)t & 0 &0 \\
 g_{13}&0 & (\beta_1+\beta_3) t & 0  \\
g_{23} &0& 0&(\beta_2+\beta_3)t  \\
\end{array} \right).
\label{h4b}
\end{equation}
It describes the four-state bowtie model, whose expression for the transition probability matrix is known exactly
 \cite{bow-tie}:
\begin{equation}
\hat{P}^{bt} = \left( \begin{array}{cccc}
(p_1p_2p_3)^2 & pq_1 & p p_1 q_2 &pp_1p_2q_3 \\
pq_1& p_1^2& p_1q_1q_2 &p_1p_2q_1q_3 \\
 pp_1q_2&p_1q_1q_2 &(1-p_1q_2)^2& p_1^2p_2q_2q_3  \\
pp_1p_2q_3 &p_1p_2q_1q_3& p_1^2p_2q_2q_3&(1-p_1p_2q_3)^2  \\
\end{array} \right),
\label{p4b}
\end{equation}
where
\be
\label{ppp}
\nonumber p_1\equiv e^{-\pi g_{12}^2/(\beta_1+\beta_2)}, \quad  p_2\equiv e^{-\pi g_{13}^2/(\beta_1+\beta_3)}, \\
 p_3\equiv e^{-\pi g_{23}^2/(\beta_2+\beta_3)}, \quad q_n=1-p_n.
\ee

We can now express transition probabilities in the model (\ref{h4b}) via transition probabilities in the model (\ref{h6b}). For example, let initial state be $|0\ra$ and we consider the probability to find the fermion $\hat{c}_1$ at the end of evolution:
\be
\la \hat{n}_1 \ra = \la 0 | \hat{c}_1^{\dg} (+\infty) \hat{c}_1(+\infty) |0\ra = |S_{14}|^2+|S_{15}|^2.
\label{con1}
\ee
On the other hand, this is the probability of that, in the four-state bowtie model, the final state is either $|12\ra$ or $|13\ra$:
\be
\la \hat{n}_1(+\infty) \ra = P_{12,0}^{bt} +P_{13,0}^{bt},
\label{con2}
\ee
where the upper index ``$bt$" means that this is the element of the transition probability matrix of the bowtie
model (\ref{h4b}). Similarly, we find
\be
\la \hat{n}_2(+\infty) \ra = P_{12,0}^{bt} +P_{23,0}^{bt} =|S_{24}|^2+|S_{15}|^2,
\label{con22}
\ee
\be
\la \hat{n}_3(+\infty) \ra = P_{13,0}^{bt} +P_{23,0}^{bt} =|S_{24}|^2+|S_{14}|^2.
\label{con23}
\ee
From Eqs.~(\ref{con1})-(\ref{con23}), we derive
\be
|S_{14}|^2=P_{13,0}^{bt}, \,\,\, |S_{15}|^2=P_{12,0}^{bt}, \,\,\, |S_{24}|^2=P_{23,0}^{bt}.
\label{con3}
\ee
Considering other initial conditions, we can reconstruct all other elements of the probability matrix in the six-state model with the Hamiltonian (\ref{h6b}):
\begin{widetext}
\begin{equation}
\hat{P} = \left( \begin{array}{cccccc}
p_1^2p_2^2& p_1^2p_2q_2q_3 & p_1p_2q_1q_3&pp_1q_2& pq_1 &0 \\
p_1^2p_2q_2q_3  &p_1^2(1-p_2q_3)^2& p_1q_1q_2& pp_1p_2q_3 & 0 & pq_1 \\
p_1p_2q_1q_3 &p_1q_1q_2 &(p-p_1)^2 & 0& pp_1p_2q_3  &pp_1q_2  \\
pp_1q_2  &pp_1p_2q_3 &0 &(p-p_1)^2 &p_1q_1q_2& p_1p_2q_1q_3 \\
pq_1  & 0 &pp_1p_2q_3  &p_1q_1q_2& p_1^2(1-p_2q_3)^2& p_1^2p_2q_2q_3  \\
0 & pq_1  &pp_1q_2 & p_1p_2q_1q_3 & p_1^2p_2q_2q_3  & p_1^2p_2^2
\end{array} \right).
\label{p6b}
\end{equation}
\end{widetext}

\section{Solvable five-state bipartite model}
\begin{figure}[!htb]
\scalebox{0.535}[0.535]{\includegraphics{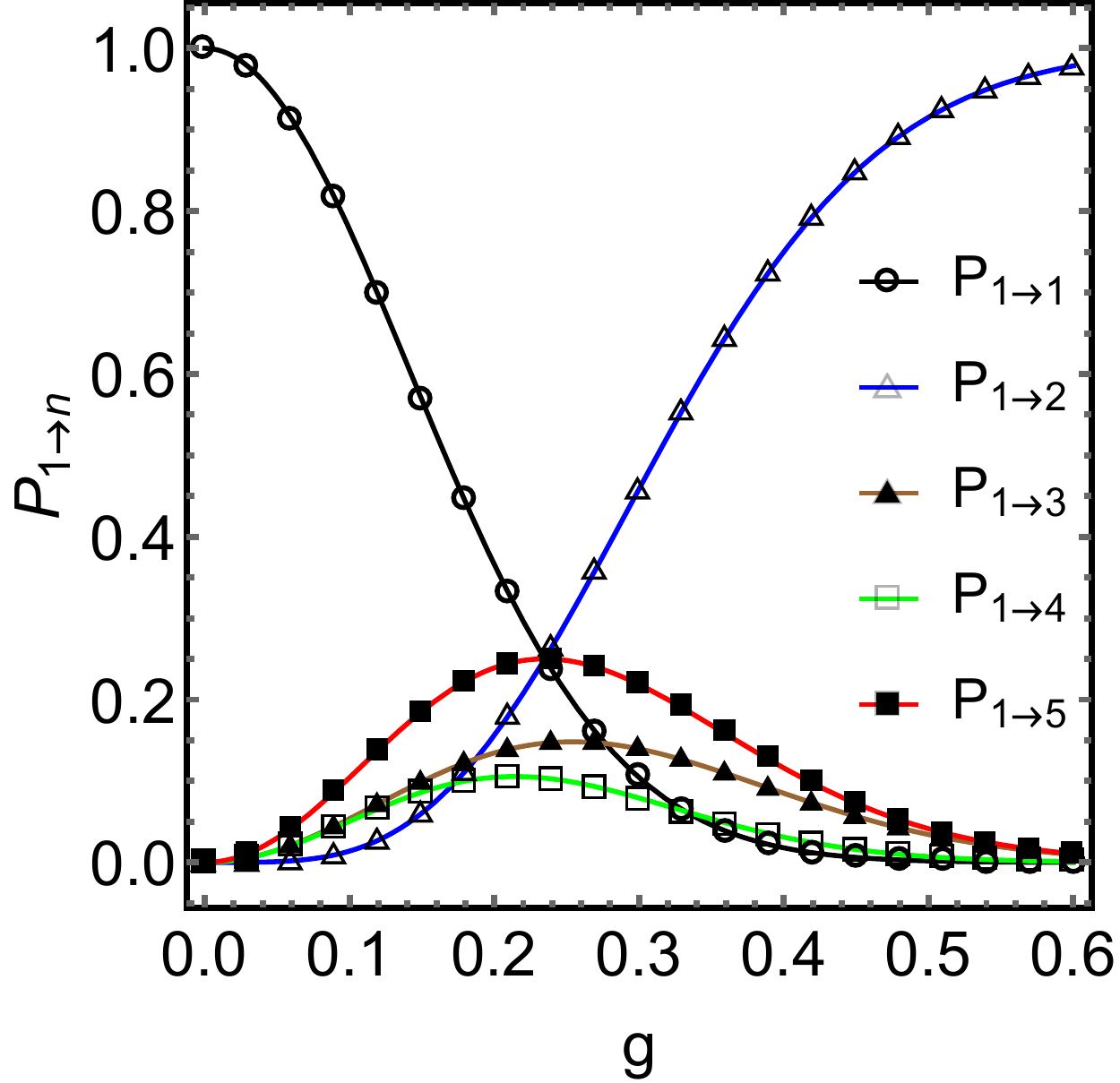}}
\caption{Transition probabilities from the level 1 to all diabatic states in the five-state model with  the Hamiltonian \eqref{5-state_Hal} and constraints (\ref{constr_5-state}) at $\tau_3=\tau_4=\tau_5=1$. Solid curves are predictions of Eq.~\eqref{P_5-state} and discrete points are results of numerically calculated $P_{1\rar n}=|U_{n1}(T,-T)|^2$, at $T=100$. The time step for simulations is $dt=0.01$. The slopes are chosen as $b = 1$, $b_3 = 4$, $b_4 = 2$, $b_5 = -3$, and the couplings are chosen to depend on a single parameter $g$ as $g_{13}=g \sqrt{b_3 - b}$, $g_{14} = g\sqrt{b_4 - b}$, $g_{15}= \sqrt{2} g \sqrt{b- b_5}$.}
\label{5-state_P1n}
\end{figure}
The general bipartite five-state model has the Hamiltonian
\begin{equation}
\hat H=\left( \begin{array}{ccccc}
b t & 0 & g_{13} & g_{14} &  g_{15}\\
0 & -b t &  g_{23} &  g_{24} &  g_{25}\\
g_{13} & g_{23} &  b_3 t  & 0 &0\\
g_{14}  & g_{24} & 0  & b_4 t  &0\\
g_{15} & g_{25} & 0 & 0   & b_5 t
\end{array}\right),
\label{5-state_Hal}
\end{equation}
where we used the fact that one can fix slopes of any two levels by gauge transformation and time-rescaling.
To find  parameters at which this model is solvable, we follow the same steps as in solution of the four-state model in Eq.~(\ref{four-h}), which was developed in \cite{Quest-LZ}. First, we add different constant parameters to the diagonal of the Hamiltonian, so that the pattern of diabatic level crossings looks like Fig.~\ref{levels-fig}(a). Next, imposing integrability conditions \cite{Quest-LZ}, we determine the constraints on parameters that make this model solvable. We then apply the semiclassical ansatz, as it is explained in \cite{Quest-LZ} in order to derive transition probabilities.
Finally, we set constant parameters at the main diagonal of the Hamiltonian to zero and obtain the model of the type (\ref{5-state_Hal}) with  parameters that correspond to the already  known transition probability matrix.

This whole procedure is straightforward but quite lengthy, so we will only list the final results.
We found that solvable models, up to redefinition of indexes, correspond to level slopes $b_3>b_4>b>0>-b>b_5$, and the couplings satisfying the following constraints:
\begin{align}\label{constr_5-state}
&\frac{g_{13}^2}{b_3-b}+\frac{g_{14}^2}{b_4-b}+\frac{g_{15}^2}{b_5-b}=0,\nn\\
&g_{2i}=\tau_i g_{1i}\sqrt\frac{b_i+b}{b_i-b},\,\,i=3,4,5,
\end{align}
where $\tau_i$ is the sign of $g_{2i}/g_{1i}$ which can be taken as either 1 or $-1$. 
 Depending on the signs of $\tau_i$'s, there are three phases in this model that correspond to different forms of transition probability matrices.
Let
\begin{align}\label{}
&p_{3}=e^{-\frac{2\pi g_{13}^2}{|b-b_3|}},\quad p_{4}=e^{-\frac{2\pi g_{14}^2}{|b-b_4|}},\nn\\
&p_{5}=p_3 p_4=e^{-\frac{2\pi g_{15}^2}{|b-b_5|}}, \quad q_k\equiv 1-p_k.
\end{align}
For $\tau_3=\tau_4=\tau_5=1$, we found
\begin{align}\label{P_5-state}
&\hat P =\left( \begin{array}{ccccc}
             p_3^2p_4^2 & q_5^2 &  p_3p_4q_3 &  p_3^2 p_4q_4 & p_3p_4q_5 \\
             q_5^2 & p_3^2p_4^2 &   p_3p_4q_3 &  p_3^2 p_4q_4 &   p_3p_4q_5 \\
              p_3p_4q_3 &  p_3p_4q_3 & p_3 & p_3q_3q_4 &  q_3q_5 \\
             p_3^2 p_4q_4   &   p_3^2 p_4q_4   &  p_3q_3q_4 & (p_4+q_3q_4)^2 &  p_3q_4q_5 \\
             p_3p_4q_5 &  p_3p_4q_5 &  q_3q_5  &   p_3q_4q_5 & p_3^2p_4^2 \\
           \end{array}\right),
\end{align}
for $\tau_3=\tau_4=-\tau_5=1$:
\begin{align}\label{}
&\hat P=\left( \begin{array}{ccccc}
p_{3}^2p_{4}^2 & 0 &  p_{3}p_{4}q_{3} & p_3^2p_4q_4 &  q_5\\
0 & p_{3}^2p_{4}^2  & q_3 &  p_3 q_4 &  p_3p_4q_5\\
p_{3}p_{4}q_{3} & q_3  &  p_3^2  & p_3q_3q_4 &0\\
p_3^2p_4q_4  &  p_3 q_4 & p_3q_3q_4 &  (p_4+q_3q_4)^2  &0\\
q_5 & p_3p_4q_5 & 0 & 0   & p_3^2p_4^2
\end{array}\right),
\label{}
\end{align}
and for $\tau_3=-\tau_4=-\tau_5=1$:
\begin{align}\label{}
&\hat P=\left( \begin{array}{ccccc}
          (p_3p_4-q_3q_4)^2 & p_3 q_4^2 &  p_3q_3 &  p_4q_4 &  p_4q_5 \\
           p_3 q_4^2 & p_3^2p_4^2 & q_3 &  p_3p_4q_4 &   p_3p_4q_5 \\
           p_3q_3  & q_3 & p_3^2 & 0 & 0 \\
           p_4q_4 &  p_3p_4q_4 & 0 & p_4^2 &  q_4q_5\\
           p_4q_5  &  p_3p_4q_5  & 0 &  q_4q_5 & p_3^2p_4^2
         \end{array}
\right).
\label{}
\end{align}
Figure~\ref{5-state_P1n} compares some of these predictions to results of our numerical simulations. The agreement is excellent.

{\it Acknowledgment}. 
This work
was carried out under the auspices of the National Nuclear
Security Administration of the U.S. Department of Energy at Los
Alamos National Laboratory under Contract No. DE-AC52-06NA25396. V.Y.C. was supported by the National Science Foundation under Grant No. CHE-1111350. N.A.S. and F.L. also thank the support from the LDRD program at LANL.

F. Li and C. Sun made equal contributions to this article.


\begin{thebibliography}{300}
\bibitem{note} In mathematical literature Stokes lines are called anti-Stokes lines and vice versa. We will accept physics nomenclature in this article.
\end{thebibliography}
\end{document}